\documentclass[]{aastex631}
\usepackage{appendix}
\usepackage{float}

\usepackage{rotating}

\usepackage{amsmath}

\usepackage{soul}
\graphicspath{{./}{figures/}}

\begin{document}

\title{The relationships among solar flare impulsiveness, energy release, and ribbon development}
\shorttitle{The impulsiveness of solar flares}
\author[0000-0002-3229-1848]{Cole A Tamburri}
\affiliation{National Solar Observatory, University of Colorado Boulder, 3665 Discovery Drive, Boulder, CO 80303, USA}
\affiliation{Department of Astrophysical and Planetary Sciences, University of Colorado Boulder, 2000 Colorado Ave, CO 80305, USA}
\affiliation{Laboratory for Atmospheric and Space Physics, University of Colorado Boulder, 3665 Discovery Drive, Boulder, CO 80303, USA.}

\author[0000-0001-8975-7605]{Maria D Kazachenko}
\affiliation{National Solar Observatory, University of Colorado Boulder, 3665 Discovery Drive, Boulder, CO 80303, USA}
\affiliation{Department of Astrophysical and Planetary Sciences, University of Colorado Boulder, 2000 Colorado Ave, CO 80305, USA}
\affiliation{Laboratory for Atmospheric and Space Physics, University of Colorado Boulder, 3665 Discovery Drive, Boulder, CO 80303, USA.}

\author[0000-0003-1325-6649]{Adam F Kowalski}
\affiliation{National Solar Observatory, University of Colorado Boulder, 3665 Discovery Drive, Boulder, CO 80303, USA}
\affiliation{Department of Astrophysical and Planetary Sciences, University of Colorado Boulder, 2000 Colorado Ave, CO 80305, USA}
\affiliation{Laboratory for Atmospheric and Space Physics, University of Colorado Boulder, 3665 Discovery Drive, Boulder, CO 80303, USA.}

\newpage
\begin{abstract}

We develop the impulsiveness index, a new classification system for solar flares using the SDO/EVE 304 \AA\ Sun-as-a-star light curves.  Impulsiveness classifies events based on the duration and intensity of the initial high-energy deposition of energy into the chromosphere.  In stellar flare U-band light curves, \cite{kowalski2013} found that impulsiveness is related to quantities such as a proxy for the Balmer jump ratio.  However, the lack of direct spatial resolution in stellar flares limits our ability to explain this phenomenon.  We calculate impulsiveness for 1368 solar flares between 04/2010 and 05/2014.  We divide events into categories of low, mid, and high impulsiveness. We find, in a sample of 480 flares, that events with high maximum reconnection rate tend to also have high impulsiveness. For six case studies, we compare impulsiveness to magnetic shear, ribbon evolution, and energy release. We find that the end of the 304 \AA\ light curve rise phase in these case studies corresponds to the cessation of PIL-parallel ribbon motion, while PIL-perpendicular motion persists afterward in most cases.  The measured guide field ratio for low and mid-impulsiveness case study flares decreases about an order of magnitude during the impulsive flare phase.  Finally, we find that, in four of the six case studies, flares with higher, more persistent shear tend to have low impulsiveness. Our study suggests that impulsiveness may be related to other properties of the impulsive phase, though more work is needed to verify this relationship and apply our findings to stellar flare physics.

\end{abstract}


\section{Introduction} \label{sec:intro}

The complex topology of the magnetic field in the Sun's upper convection zone and lower atmosphere causes interaction between different magnetic topological domains, leading to magnetic reconnection.  During reconnection, field lines belonging to different magnetic topological domains diffuse through a small-scale region, the reconnection current sheet (RCS), and merge to form new field lines in the direction perpendicular to plasma inflow.  In some cases, a solar flare results.  Reconnection in large flares initiates the process traditionally outlined with the Carmichael, Sturrock, Hirayama, Kopp, and Pneuman (CSHKP) model of a two-ribbon flare \citep{carmichael_1963,sturrock_1966,hirayama_1974,kopp_pneuman_1976}.  Large amounts of magnetic free energy are released across the electromagnetic spectrum, including in the ultraviolet, X-ray, and occasionally white light.  

After magnetic reconnection commences, precipitation of particles from the coronal reconnection site and interaction with ambient chromospheric plasma releases energy in ultraviolet/extreme ultraviolet (UV/EUV) and hard X-ray (HXR) wavelengths. We observe the sites of energy deposition in the chromosphere as flare ribbons in the EUV and UV. Flare ribbons, typically observed in He {\tiny II} 304 \AA\ and 1600 \AA, are located at the intersection of separatrices or quasi-separatrix layers with the chromosphere \citep[e.g.][]{savcheva2015,savcheva2016}.  There, the energy of precipitating particles is transferred to chromospheric plasma \citep[e.g.][]{aschwanden2002}.  He {\tiny II} emits energy at 304 \AA\ at temperatures of 50,000 K, and the 1600 \AA\ line represents the continuum at about 10,000 K with a strong contribution from C {\tiny IV}.   HXR emission is thought to be thick-target non-thermal brehmsstrahlung radiation due to the interaction of precipitating particles with ambient plasma, with photon energies having effective temperatures of $10^8$ K. 

\begin{figure}
\begin{center}
   \includegraphics[width=.6\linewidth]{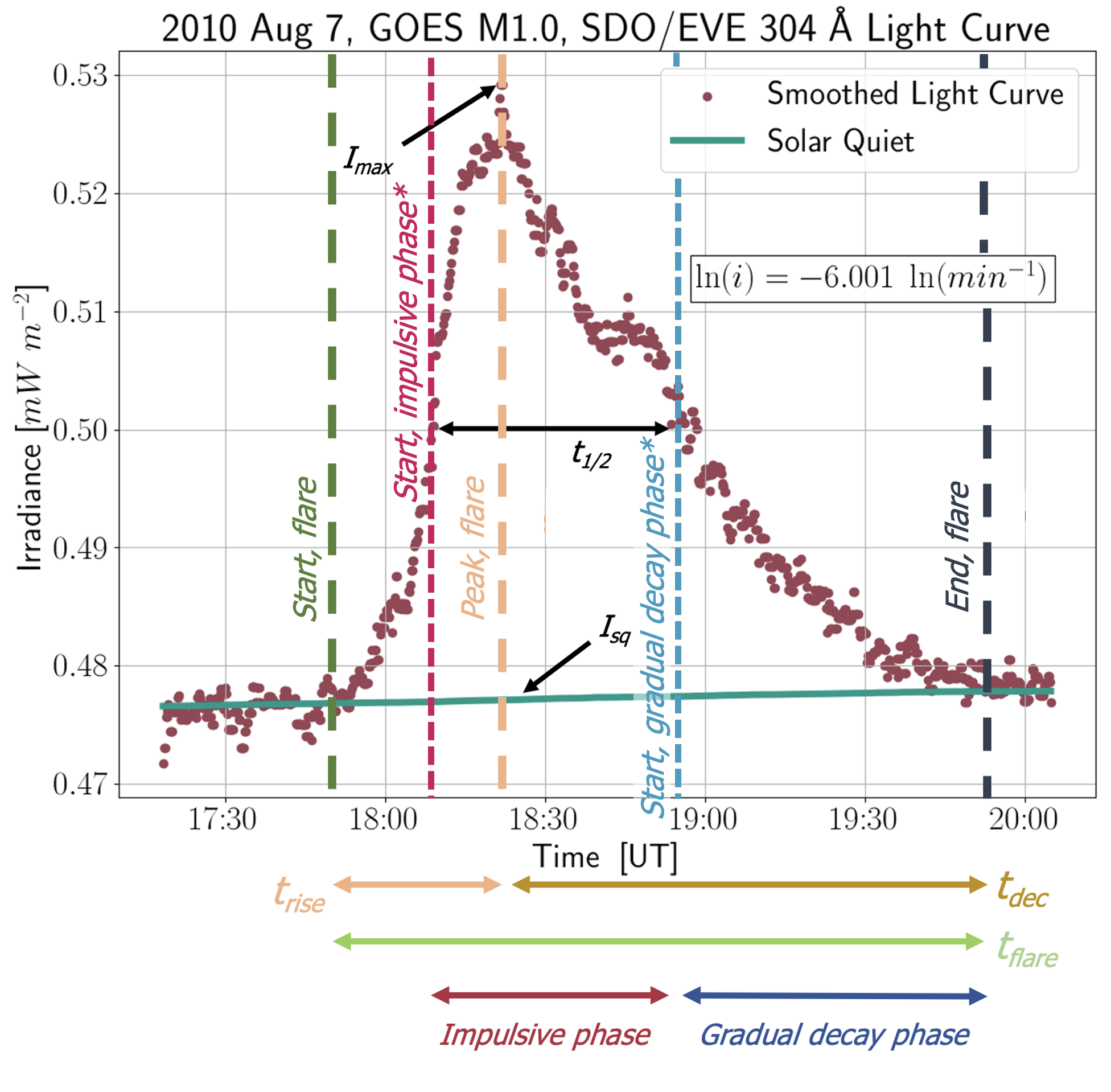}
   \caption{A sample SDO/EVE 304 \AA\ light curve from the GOES class M1.0 flare on 7 August 2010, with the relevant parameters for determination of the impulsiveness index identified: the maximum irradiance $I_{max}$, the estimated non-flare irradiance $I_{sq}$, and the temporal full width at half height $t_{1/2}$.  The impulsiveness of this event is $\ln(i) = -6.001 \ln(min^{-1}$), in the category of ``low impulsiveness."  The flare start time, impulsive phase start time, flare peak time, gradual decay phase start time, and flare end time are labeled with green, maroon, tan, blue, and black vertical dotted lines, respectively.  Note that the impulsive and gradual decay phase start times are approximate.  Below the figure, we show the periods associated with $t_{rise}$, $t_{dec}$, $t_{flare}$, the impulsive phase, and the gradual phase, as are referenced throughout our work. A color version of this figure is available online.}
  \label{examplecurve}
  \end{center}
\end{figure}

HXR and EUV light curves show a sharp peak in emission during chromospheric energy deposition.  This peak occurs during the ``impulsive" flare phase, which can last for a period as short as tens of seconds, or, in some flares, as long as several minutes.  The impulsive phase includes a fast rise, peak, and fast decay \citep{kowalski2013} and is followed by the ``gradual" (or ``gradual decay") phase, characterized by a slower, exponential decay of the light curve \citep{hawley_2014,kowalski2013}, during which time UV/EUV emission returns to solar quiet \citep{hawley1995}. As an initial choice, for the reasons discussed in Section \ref{sec:datainstr}, we use the temporal full width at half height in the 304 \AA\ line as a proxy for the impulsive phase.  In Section \ref{sec:conc}, we discuss how the upper atmospheric sources of the 304 \AA\ line in the flaring solar atmosphere may motivate a larger study to determine whether there is another option to use for the development of the impulsiveness index. \footnote{N.B. The distinction between the ``impulsive" and ``gradual" phases (see Figure \ref{examplecurve}) is not the same as that between the ``rise" and ``decay" phases ($t_{rise}$ and $t_{dec}$ in Figure \ref{examplecurve}).  While the rise phase, in our work, is the period from the start of the flare to the peak intensity and the decay phase is the period from the peak of flare intensity to the return to solar quiet (both measured in the 304 \AA\ light curve), the impulsive phase is traditionally defined as the period of intense HXR emission directly before and after the peak HXR intensity, while the gradual decay phase begins after the majority of HXR emission has occurred.  In the EUV, the gradual phase begins some time after the beginning of the decay of the light curve, until the return to solar quiet. We discuss issues with characterizing the impulsive phase based on 304 \AA\ observations in Section \ref{sec:conc}.} 

To better understand solar flares, past work has recreated spectral and spatial flare observations with detailed models and simulations.  Two-dimensional models for reconnection \citep[e.g. Section 4.1.1 in][]{shibata_magara_2011} and one-dimensional simulations of chromospheric loop heating \citep[e.g.][]{allred2015,kowalski2017} have proven instrumental in reconciling solar flare physics with observations.  However, recent work has revealed complex phenomena involved in three-dimensional magnetic field simulations \citep[e.g.][]{dahlin2015,dahlin2016,arnold2021,daldorff2022}.  The paradigm shift from antiparallel (2-dimensional) to component (3-dimensional) reconnection \citep[e.g.][]{dahlin2014}, in which reconnecting magnetic field lines are not antiparallel in 3D space, further complicates the picture of reconnection.  Component reconnection involves the presence of a guide field component of the magnetic field in the direction along the RCS.  

In both simulations and observations, the magnitude of the guide field is related to the rate and timing of energy release.  A strong guide field is associated with a large amount of magnetic shear, characterized by a strong guide field relative to the reconnecting component of the magnetic field, which is perpendicular to the RCS.  A more sheared magnetic field indicates that the field is in a non-potential state. The presence of a non-potential field is necessary for the storage of the free energy which is released in other forms as the magnetic field relaxes to a more potential state \cite[][and references therein]{tiwari_2010,janvier_2015,leake_2022}.  The relaxation of the magnetic field is observed in the flare ribbons as a ``strong-to-weak” shear pattern \citep{alexander_2001,aulanier_janvier_schmieder_2012,kitahara_kurokawa_1990,machado_1985,su_golub_van_ballegooijen_2007,su_golub_van_ballegooijen_gros_2006}.  

Observational studies such as ours use flare ribbons to measure the evolution of magnetic shear. \cite{zhou_2009} present evidence that tracking ribbon shear via the relative position of the positive and negative polarity ribbons is an appropriate method for quantification of magnetic shear, as in e.g. \cite{qiu2010} and \cite{qiu2022}. \cite{qiu_2023} compute shear for a two-ribbon GOES class M6.9 flare occurring on 2014 December 18 (1) using EUV images of the post-reconnection flare loops (PFRLs) and (2) via analysis of the footpoints in 1600 \AA\ UV images.  They find that a similar strong-to-weak shear pattern is present in both versions of calculated shear, although it is difficult to determine the extent to which a one-to-one relationship between the orientation of PFRLs and flare ribbon fronts exists from analysis of a single flare.

Despite the proportionality between the amount of pre-flare stored energy and the intensity of energy release, the early presence of a strong guide field suppresses particle acceleration during a flare \citep{dahlin2021,arnold2021}.   Shear reduction leads to intense particle acceleration through the increased relative efficiency of Fermi acceleration \citep{dahlin2016, dahlin_2020, dahlin2021,qiu2010,litvinenko_1996,arnold2021}.  \cite{dahlin2016} discuss this phenomenon, also simulated by \cite{arnold2021}, and show that in cases of weak guide field, Fermi reflection occurs efficiently over macroscale sizes, producing large numbers of nonthermal electrons.  

The change in the total kinetic energy of an electron \citep{dahlin2016} can be written as

\begin{equation}\label{eq:elecen}
\frac{d\epsilon}{dt}=qE_{\parallel}v_{\parallel}+\frac{\mu}{\gamma}\left(\frac{\partial B}{\partial t} + \mathbf{u}_E\cdot\nabla B \right)+\gamma m_e v^2_{\parallel}(\mathbf{u}_E\cdot\mathbf{\kappa}),
\end{equation}

\noindent where $\epsilon = (\gamma - 1)m_e c^2$ is the kinetic energy, $q$ is the electron charge, $E_{\parallel}$ is the magnitude of the electric field parallel to the local magnetic field $\mathbf{B}$, $v_{\parallel}$ is the magnitude of the electron velocity component parallel to the magnetic field,\footnote{$v_{\parallel}$ is not the same quantity as $v_{\parallel,\pm}$ in Figure \ref{fig:ribbonev}, which refers to the PIL-parallel ribbon motion.} $\mu$ is the magnetic moment, $\gamma$ is the relativistic Lorentz factor, $\mathbf{u}_E = c\mathbf{E}\times\mathbf{B}/B^2$ is the drift velocity, $m_e$ is the electron mass, and $\mathbf{\kappa}$ is the magnetic curvature.

In Equation \ref{eq:elecen}, the first term on the right-hand side corresponds to acceleration due to the electric field parallel to the local magnetic field.  The second term captures betatron acceleration, which has a cooling effect on the perpendicular component of the electron energy \citep{dahlin2016,dahlin_2020}.  The third term accounts for acceleration due to Fermi reflection in a magnetic mirror.  The guide field, which does not appear in Equation \ref{eq:elecen}, attenuates Fermi acceleration, as described in \cite{dahlin2016}, both by decreasing the frequency of Fermi reflections and decreasing the energy gain for each reflection. 

As a result, when the guide field is strong, the first term in Equation \ref{eq:elecen} dominates the heating of electrons in the direction parallel to the magnetic field.  Energization due to $E_\parallel$ is weaker than energization due to Fermi acceleration.  Therefore, when the former is the dominant acceleration mechanism because of the presence of a strong guide field, electron acceleration is relatively weak \citep{dahlin2016,arnold2021}.  If, however, the guide field is weak relative to the reconnecting component of the magnetic field, Fermi acceleration becomes the most efficient electron acceleration mechanism in the direction parallel to the magnetic field, and more significant nonthermal electron acceleration occurs.

Other studies have analyzed the impact of the guide field on flare development.  \cite{daldorff2022} investigate the effect of guide field strength on reconnection rate by simulating a 3D RCS. They find that a strong guide field slows the rate of reconnection through the interaction of the oblique modes of the vector potential perturbation, which have associated flux ropes that are misaligned with the guide field. By measuring magnetic reconnection rate, comparing different means of quantifying magnetic shear, and comparing to non-thermal electron flux for a case study flare, \cite{qiu_2023} provide further evidence that ``intermediate" shear values may facilitate efficient particle acceleration.

Not all flares demonstrate the strong-to-weak shear pattern.  For example, the ``hook" structure observed by \cite{aulanier_dudik_2019} and discussed by \cite{zemanova_2019} can occur at the feet of an erupting flux rope connecting the opposite ends of conjugate ribbons.  The simulations of \cite{dahlin2021} also display this behavior.  Such a configuration may suggest an increasing value for shear when derived from the ribbon areas in the ``hook" structure.  \cite{qiu2022} also discuss this complication.  Therefore, it is difficult to find a universal connection between magnetic configuration (as inferred from flare ribbon morphology) and flare energy release.  We return to this discussion in Section \ref{sheardisc}.

In this paper we introduce the impulsiveness index \textit{i} in order to substantiate the simulation results from \cite{daldorff2022} and \cite{dahlin2021} and, more generally, to classify flares in a manner which effectively reflects the rate and magnitude of energy release in the impulsive phase.  We define $i$, below, as the ratio of peak normalized flare irradiance to flare impulsive phase duration in the 304 \AA\ light curves, measured by the Solar Dynamics Observatory Extreme Ultraviolet Experiment (SDO/EVE), following \cite{kowalski2013}:

\begin{equation}\label{eq:imp}
    i = \frac{[(I_{max}-I_{sq})/I_{sq}]}{t_{1/2}},
\end{equation}

\noindent  where $I_{max}$ is the peak irradiance in the 304 \AA\ line, $I_{sq}$ is the estimated solar quiet at the time of the peak flare intensity in the 304 \AA\ line, and $t_{1/2}$ is the full width at half-maximum of the light curve relative to the quiet sun irradiance, as in Figure \ref{examplecurve}.  In a classification scheme based on impulsiveness, flares with a low peak intensity may be grouped together with those of a high peak intensity, provided that the former have an appropriately short overall duration.

Measured from stellar, or Sun-as-a-star, flare light curves, the impulsiveness index captures the rate and timing of energy release. This index is empirically related to spectral properties which are not reflected in peak GOES X-ray flux \citep{kowalski2013,kowalski_2016}. 
 \cite{kowalski2013} originally developed the impulsiveness index for a selection of M dwarf stellar flares.  They found that impulsiveness is anti-correlated with a proxy for the Balmer jump ratio ($\chi_{flare,peak}$, which measures the difference in continuum flux on either side of the Balmer series limit), with the ratio of H$\gamma$ line flux to the continuum intensity at 4170 \AA\ (H$\gamma$/C4170), and with the ratio of Balmer line flux to total flux in the 3420 - 5200 \AA\ band.  As shown in Figure 21 of \cite{kowalski2013}, more impulsive flares have a stronger continuum flux relative to the H$\gamma$ line during both the peak and gradual phases of stellar flares.  Since impulsiveness is related to the intensity of flare-related emission relative to blackbody-like emission in \cite{kowalski2013}, the index may have a stronger relationship with plasma temperature, optical depths, and the development of flare ribbons than other flare classification schemes do. 

While high-cadence point-source photometry of stellar flares \citep[e.g.][]{kowalski_2016,kowalski_2023} has allowed for the discovery of the empirical relationship between impulsiveness and spectral stellar flare properties, the lack of direct spatial resolution of stellar flare observations obfuscates the physical explanation for such relations.  We attempt to provide physical insight by turning to observations of the Sun and beginning an investigation into connections between impulsiveness and measured properties of solar flares.  It is thought that there is a connection between the consequences of electron acceleration and large-scale flare evolution \citep{holman2011}.  Therefore, we look at a readily constrained quantity, the reconnection rate, which is generally accepted to be a proxy of particle acceleration and is derived from large-scale evolution.  In Section \ref{sec:statstudy} we compare impulsiveness to a number of solar flare quantities and find little correlation besides a relationship between impulsiveness and peak reconnection rate. In Section \ref{sec:casestudy}, an analysis of six individual flares allows a closer examination of the magnetic topology evolution and the relationship with spatially integrated light curve properties such as impulsiveness.   

Because of the variety and complexity of flares, it is difficult to find clear, physically meaningful correlations between different flare parameters \cite[e.g.][]{kazachenko2017,harra_2016}.
However, previous work and our interpretation of impulsiveness provide clues for what we expect the relationship to be. EUV emission occurs during the flare impulsive phase as a consequence of particle precipitation, and the rates of particle precipitation are related to the magnitude of magnetic shear \citep{dahlin_2020,arnold2021}. Therefore, the impulsiveness index, which is derived from EUV light curves indicating emission in flare ribbons, may also be related to the storage of magnetic free energy in the form of shear in some flares.

If there is a relationship between impulsiveness and the timing and rate of particle precipitation, chromospheric energy release, and coronal loop orientation, we should see the relationship in a sample of case study flares.   For six events, we track the motion of chromospheric ribbons using 1600 \AA\ images from the Solar Dynamics Observatory Atmospheric Imaging Assembly (SDO/AIA). We analyze the geometrical properties of chromospheric flare ribbons relative to the polarity inversion line (PIL), which separates flare ribbon regions of opposite magnetic polarity.  We use EUV and HXR emission to characterize the motion and energy release associated with flare ribbons and particle precipitation.  Our analysis reveals distinct PIL-parallel (elongation) and PIL-perpendicular (separation) components of ribbon motion.  We find, for the six flares analyzed, that PIL-parallel ribbon motion occurs mainly in the rise phase, while PIL-perpendicular ribbon motion persists longer for most of the case study flares.  Via decomposition of ribbon motion, we quantify magnetic shear and the reconnection component of the electric field in the RCS.  We compare these quantities to the development of EUV and HXR light curves, focusing in particular on the range of impulsiveness values represented by the case studies.  We discuss case study results in the context of the hypothesis that more impulsive events may experience an earlier, steeper decline in magnetic shear.

This paper is organized as follows.  In Section 2, we describe observations and the \verb+RibbonDB+ catalog.  In Section 3, we describe the impulsiveness index $i$ and present a statistical analysis of the index, including a comparison to other flare quantities.  In Section 4, we describe our methods for analysis of flare ribbons and present six case studies with a range of impulsiveness values.  Finally, in Section 5, we summarize our main findings and suggest possible avenues for application of our work.

\section{Data}\label{sec:data}

In this section, we describe the data sources (Section \ref{sec:datainstr}) and the catalog of flare events (Section \ref{RibbonDB}) used in our study.

\subsection{Observations}\label{sec:datainstr}

We use the SDO/EVE Multiple EUV Grating Spectrographs-A (MEGS-A) 304 \AA\ Sun-as-a-star data \citep{woods2010} to calculate the impulsiveness index.  MEGS-A is a grazing-incidence spectrograph observing roughly the 60 to 370 \AA\ range. We use the 304 \AA\ line due to (1) sensitivity of this line to the lower layers of a flaring solar atmosphere, including a significant contribution from chromospheric flare ribbons; (2) the availability of 304 \AA\ SDO/EVE data, intersecting the \verb+RibbonDB+ catalog \citep{kazachenko2017} and the first years of solar cycle 24; (3) the lack of saturation effects, as, for example, occurs in 1600 \AA\ images; and (4) the 10-second time cadence of observations. Since the impulsive flare phase can occur on scales as fast as tens of seconds, the high-cadence light curves are necessary. Despite this justification for the use of the 304 \AA\ line, it is possible that another line or bandpass would be a better option to develop the impulsiveness index.  We have included a discussion of this topic in Section \ref{sec:conc}.

We identify flares and process the data using the \verb+RibbonDB+ catalog, described in Section \ref{RibbonDB}. We then calculate the impulsiveness index for 1368 flares found in \verb+RibbonDB+ between 2010 April 30 and 2014 May 26.  On 2014 May 26, MEGS-A experienced an anomaly, making flare observations of the 304 \AA\ line after this date impossible.   

Saturation is a major issue in solar flare imaging, and is widely present in SDO/AIA 1600 \AA\ images, affecting at least $10^5$ image frames \citep{guastavino2019}.  Desaturation, as for example in \cite{kazachenko2017}, relies on non-saturated images before and after the image frame containing saturated pixels, but could neglect the key geometrical evolution of flare ribbons during the impulsive phase.  To avoid saturation altogether in this study, we only consider flares of GOES class C or M for case study analysis, as these are less likely than X-class events to suffer from saturation. While developing the impulsiveness index, however, we consider all flare classes above GOES C1.0.

To identify the magnetic polarity of flare ribbons, we use the normal component of the magnetic field, $B_n$, from the SDO Helioseismic and Magnetic Imager (SDO/HMI) vector magnetograms.  SDO/HMI makes use of the Zeeman effect to compute Stokes parameters and measure the vector magnetic field, among other data products, with the Fe {\tiny I} 6173 \AA\ absorption line \citep{scherrer2012,schou2012}. Since the structure of the photospheric magnetic field in the active region area does not change appreciably over the timescale corresponding to a flare, we require only the image from SDO/HMI immediately prior to the flare start time for each event analyzed.  

We use the Object Spectral Executive (OSPEX) software to access Fermi Gamma-ray Burst Monitor (Fermi/GBM) data \citep{meegan_2009}.  Fermi/GBM uses 12 NaI and 2 BGO scintillation detectors to measure the flux of gamma-rays in the range of 8 keV to 40 MeV \citep{wilson-hodge_2012} and has been employed in many flare studies \citep[e.g.][]{inglis2015,inglis2016}.  We use flux in the range of 25 - 300 keV to quantify HXR flux.

\subsection{The \texttt{RibbonDB} catalog}\label{RibbonDB}
\verb+RibbonDB+ is a catalog of 3137 solar flares taking place between April 2010 and April 2016 of GOES class C1.0 or higher, within 45$^o$ of the central meridian \citep{kazachenko2017}.  For our analysis, we consider 2049 events between 2010 April 30 and 2014 May 26, when the SDO/EVE MEGS-A device was operational \citep{machado2018}.    

The \verb+RibbonDB+ catalog provides reconnection fluxes for each flare in the sample.  To estimate these fluxes, \verb+RibbonDB+ uses instantaneous ribbon masks, $N_{cut}$, which are regions where the 1600 \AA\ emission exceeds the median image intensity multiplied by a factor $c$ (where $c$ = 8 in our analysis).  We define the unsigned reconnection flux $\Phi_{ribbon}$ as in Equation \ref{eq:recflux}: 

\begin{equation}
    \Phi_{ribbon}(t_k) = \int_{I_c}|B_n(t_k)|dS(t_k) 
    \approx \sum_{i,j}|B_n(x_i,y_j,t_{\mathit{HMI}})|M_{cut}(x_i,y_j,t_k)ds_{ij}^2,
    \label{eq:recflux}
\end{equation}

\noindent where $B_n$ is the normal component of the magnetic field measured at the time of the SDO/HMI image ($t_{\mathit{HMI}}$), and $dS(t_k)$ is the ribbon area above the intensity cutoff $c$ at time $t_k$, approximated discretely by the cumulative ribbon mask $M_{cut}(x_i,y_j,t_k)$ times the pixel area $ds^2_{ij}$.  The cumulative ribbon mask at time $t_k$ includes all brightened pixels (with a value of one) above the cutoff criteria in all previous instantaneous ribbon masks $N_{cut}(x_i,y_j,t_k)$, which include only newly-brightened pixels. The cumulative and instantaneous ribbon masks are related by $M_{cut}(x_i,y_i,t_k) = M_{cut}(x_i,y_j,t_{k-1})\cup N_{cut}(x_i,y_j,t_{k})$.  Here we use the approach from \cite{kazachenko2017} except that we do not consider intensity cutoffs other than $c = 8$. Pixels below the intensity cutoff are set to zero in the ribbon masks. 

The \verb+RibbonDB+ catalog also provides the peak reconnection flux rates for flares in the catalog.  We define the reconnection rate $\dot{\Phi}_{ribbon}$ at time $t_k$ as the time derivative of the reconnection flux:  

\begin{equation}
\dot{\Phi}_{ribbon}(t_k) = \frac{\partial\Phi_{ribbon}(t_k)}{\partial t}
    \label{eq:recrate}
\end{equation}

\noindent where variables are defined as in Equation \ref{eq:recflux}.  In Section \ref{sec:statstudy} we compare impulsiveness to the peak reconnection flux rate. 
 We define the peak reconnection flux rate $\dot{\Phi}_{ribbon,peak}$ as the average of the magnitudes of the positive and negative reconnection rates in order to reduce error:
 
 \begin{equation}
\dot{\Phi}_{ribbon,peak} = \frac{|\dot{\Phi}_{ribbon,peak}^+|+|\dot{\Phi}_{ribbon,peak}^-|}{2}
    \label{eq:recratepeak}
\end{equation}

  \noindent where positive and negative values are signed peak reconnection rates in each polarity. Since the positive and negative reconnection rates should theoretically be equal throughout the flare, we can estimate the error in peak reconnection rate values using the imbalance between positive and negative reconnection rates as defined in \cite{kazachenko2023}:

\begin{equation}
\dot{\Phi}_{ribbon,imb} = |\;|\dot{\Phi}_{ribbon,peak}^+| - |\dot{\Phi}_{ribbon,peak}^-|\;|
    \label{eq:imb}
\end{equation}

\noindent where variables are defined as in Equation \ref{eq:recratepeak}.  We evaluate the relationship between impulsiveness and other flare properties using Pearson's correlation coefficient $r^2$, which measures the strength of the linear relationship between two variables, and the Spearman rank correlation coefficient $r_s$, which measures the strength of the potentially non-linear monotonic relationship between two variables.  For these metrics, values of $ r^2,\; |r_s| < \pm 0.3$ suggests that the variables being compared have very little, if any, correlation.  $\pm 0.3 < r^2,\; |r_s| < \pm 0.6$ suggests that the variables are moderately correlated, and $\pm 0.6 < r^2,\; |r_s| < \pm 1$ suggests that the variables are strongly correlated. 

\section{The impulsiveness index}\label{sec:statstudy}

\subsection{Definition}
We define the impulsiveness index \textit{i} as:

\begin{equation}\label{eq:imp2}
    i = \frac{[(I_{max}-I_{sq})/I_{sq}]}{t_{1/2}},
\end{equation}

\noindent  where $I_{max}$ is the peak irradiance $[mW \; m^{-2}]$ in the 304 \AA\ line, $I_{sq}$ is the estimated solar quiet at all times during the flare, and $t_{1/2}$ is the full width at half-maximum of the light curve relative to the quiet sun irradiance. 
$t_{1/2}$ is an approximation of the impulsive phase duration, the period of energy deposition into the lower solar atmosphere from the coronal reconnection site.  $I_{sq}$ is estimated by smoothing the 304 \AA\ light curve with a large window, thereby removing flare signatures and providing an estimate for the ``non-flare" intensity which does not depend solely on the pre-flare or post-flare intensity.  By employing the smoothing method for solar quiet estimation, we minimize the contribution from recent flares to the estimated solar quiet intensity.  While there will always be some contribution to the non-flare intensity from other active regions or previous flares, particularly when using Sun-as-a-star light curves, the smoothing technique reduces the impact of flare signatures on the estimated solar quiet.

\cite{kowalski2013} originally developed the index using stellar flare data in the optical U-band ($\lambda$ $\sim$ 3250 - 3950 \AA).  While U-band observations are appropriate for the study of dMe flares, observations in the EUV (here, the SDO/EVE 304 \AA\ flux) exhibit chromospheric energy release effectively, for a much larger number of flares than in the optical, and with sufficiently short time resolution to capture short-duration events.

We determine the impulsiveness index for all flares in the \verb+RibbonDB+ catalog during which SDO/EVE MEGS-A was in operation.  Flares occur between 2010 April 30 and 2014 May 26.  In order to apply the index to as large a sample of flares as possible, we apply a semi-autonomous method of flare parameter determination as described in Section \ref{lcanalysis}.

Not all flare data in the \verb+RibbonDB+ catalog are of sufficient quality to include in an initial study of impulsiveness.  As described in Appendix \ref{lcanalysis}, we are unable to process some events to the degree of confidence required to compare impulsiveness to other flare properties.  Therefore, we have developed a program to confidently identify flare temporal parameters (start, peak, and end times), determine flare parameters relevant to impulsiveness ($I_{max}$, $I_{sq}$, and $t_{1/2}$), and vet the flares for clarity of signal.  The result is a selection of ``best-performing" 480 flares (1) which include representation from low, mid, and high impulsiveness categories, and (2) for which we are confident in the accuracy of the impulsiveness index calculation.  We describe the processing, parameter identification, and vetting of events for the statistical study in Appendix \ref{sec:appendix}.

\subsection{Relationship of impulsiveness to other physical quantities}\label{imprel}

\begin{figure}
   \includegraphics[width=\textwidth]{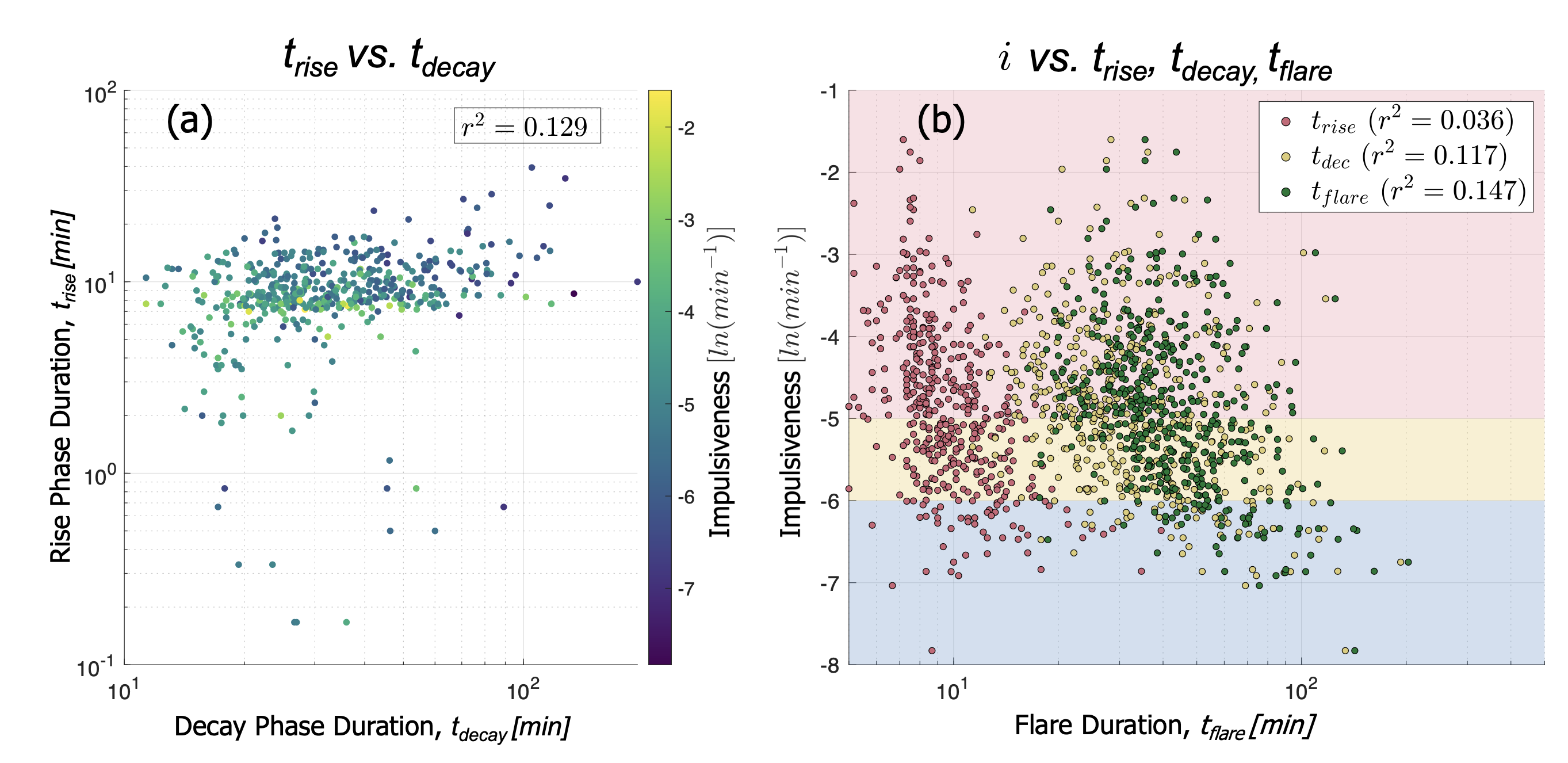}
    \caption{(a) The relationship between rise and decay phase duration for the sample of best-performing 480 flares.  While there is a slight linear relationship, there is clearly a large amount of scatter, suggesting that neither duration can be predicted by the other. The Pearson's correlation coefficient for the linear fit is low, at $r^2 = 0.129$.  Events with low rise and decay phase duration tend to be more impulsive. (b) A comparison of impulsiveness to rise phase duration (red), decay phase duration (yellow), and overall flare duration (green) for the best-performing 480 flares.  Blue, yellow, and red shaded areas indicate regions of low, mid, and high impulsiveness respectively.}
  \label{impdur}

\end{figure}

We compare impulsiveness to other flare properties for the selection of 480 best-performing light curves as described above.  First, in Figure \ref{impdur}(a), we compare the rise- and decay-phase durations ($t_{rise}$ and $t_{dec}$ respectively, where these parameters are defined in Figure \ref{examplecurve}, based on the 304 \AA\ light curve) of flares in the sample.  We find that there is a low Pearson's correlation coefficient of $r^2 = 0.129$ between decay phase duration and rise phase duration. 
Therefore rise phase duration is not a predictor of decay phase duration, and vice versa. \cite{hawley_2014} came to a similar conclusion for a sample of nearly 1000 white-light stellar flares in Kepler data from the star GJ 1243 of spectral type M4.

In Figure \ref{impdur}(b), we compare impulsiveness to rise phase duration, decay phase duration, and overall flare duration. Neither the rise phase duration nor decay phase duration correlate with the impulsiveness index ($r^2 = 0.036$ and $r^2 = 0.117$, respectively).  Total flare duration has the strongest correlation with impulsiveness ($r^2 = 0.147$).  The slope of this relationship is -0.894 with a 95\% confidence range of [-1.087,-0.700].  

Given the formulation of impulsiveness (i.e. the inverse proportionality with $t_{1/2}$), we would expect the log-log plot comparing impulsiveness to flare duration (from the flare start time to the flare end time, as indicated in Figure \ref{examplecurve}) to have a slope near -1 if the flare duration is linearly correlated with $t_{1/2}$, our proxy for impulsive phase duration.  We note, however, that impulsive phase duration and flare duration are not directly correlated. 
For the flares studied, there is a weak positive linear correlation (not shown) between $t_{flare}$ and $t_{1/2}$, with a correlation coefficient of $r^2 = 0.55$. This result suggests that flare impulsive phase duration and overall duration are representative of different physical processes, just as in \cite{hawley_2014} overall flare duration is not related to rise phase duration in a survey of flares from \textit{Kepler} short-cadence M dwarf observations. 

\begin{figure}
\begin{center}
   \includegraphics[width=.8\linewidth]{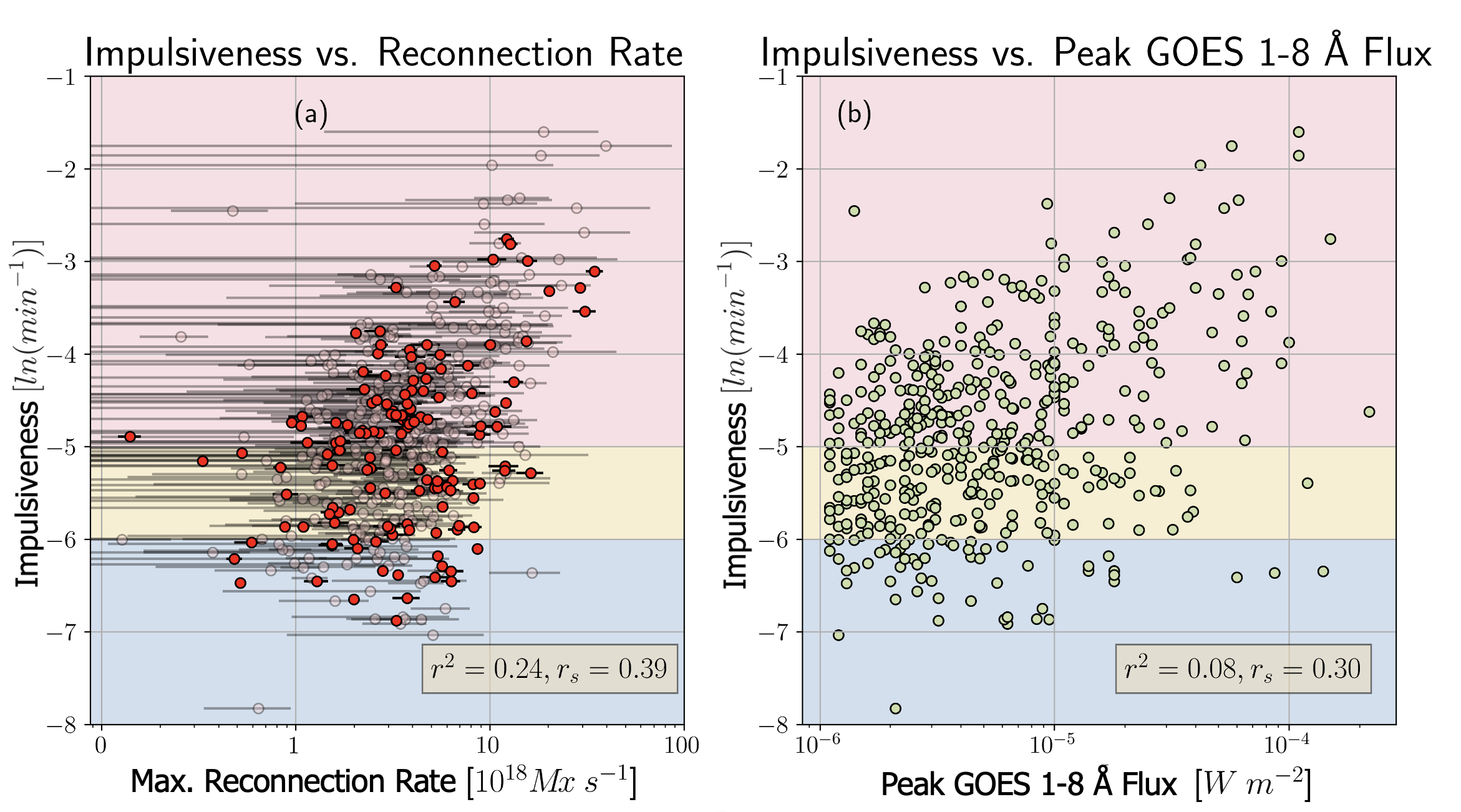}
   \caption{(a) Comparison of impulsiveness index with peak reconnection rate.  The linear correlation is weak ($r^2=0.24$) but the variables have a moderately strong monotonic, non-linear relationship ($r_s = 0.39$). Events with small error in the reconnection flux rate, for which the imbalance $\dot{\Phi}_{ribbon,imb}$ between positive and negative values is less than 20\% of the maximum reconnection rate, are opaque and shown in red.  All others are pink and transparent.  (b) Comparison of impulsiveness index with peak GOES 1-8 \AA\ soft X-ray flux.  The correlation is weak ($r^2=0.08$) but the monotonic, non-linear relationship is moderately strong ($r_s = 0.30$).  Blue, yellow, and red shaded regions indicate low, mid, and high impulsiveness respectively. Error bars indicate the imbalance $\dot{\Phi}_{ribbon,imb}$ between positive and negative reconnection rate.}
  \label{imp_goes_recrate}
  \end{center}
\end{figure}

In Figure \ref{imp_goes_recrate}(a) we compare impulsiveness to the peak reconnection rate $\dot{\Phi}_{ribbon,peak}$, defined in Equation \ref{eq:recratepeak}. We also compare impulsiveness to the reconnection flux (not shown), and in Figure \ref{imp_goes_recrate}(b), the peak GOES 1-8 \AA\ soft X-ray flux. There is a weak linear relationship between impulsiveness and peak reconnection rate (Pearson's correlation coefficient $r^2=0.24$). The Spearman rank correlation coefficient of $r_s=0.39$ suggests a moderate monotonic relationship between impulsiveness and peak reconnection rate. Also, of the 22 events with reconnection rate above $15\times10^{8} Mx \: s^{-1}$, 19 (86\%) are in the ``high-impulsiveness" category ($\ln[i] > -5.0 \ln[min^{-1}]$).  The number of events from the sample of 480 which are ``high-impulsiveness" is 261 (54\%), indicating that events with high reconnection rate are highly impulsive relative to the population of all events. 
Following \cite{kazachenko2023}, if we exclude events for which the imbalance in the peak reconnection rate $\dot{\Phi}_{ribbon,imb}$ defined in Equation \ref{eq:imb} is greater than 20\% of the averaged peak reconnection rate value, we are left with 140 events (red markers in Figure \ref{imp_goes_recrate}(a)).  For those 140 events, the Spearman rank correlation coefficient is $r_s = 0.29$, and 14 out of 17 events with high peak reconnection rate are also highly impulsive ($\ln[i] > -5.0 \ln[min^{-1}]$). 

Interestingly, as shown in Figure \ref{imp_goes_recrate}(b), there is a weak linear relationship between impulsiveness and peak GOES soft X-ray flux (1-8 \AA), with a correlation coefficient of $r^2 = 0.08$, although the Spearman rank correlation coefficient lies barely in the ``moderate" correlation category, at $r_s = 0.30$.  The lack of a strong correlation between impulsiveness and peak GOES SXR flux emphasizes that the impulsiveness index may capture a different quantity than the flare classification system based on peak GOES SXR flux.  Peak intensity in the 304 \AA\ line largely captures the intensity of energy deposition in the chromosphere, while the duration of the impulsive phase reflects the time period corresponding to chromospheric energy release.  The magnitude of peak GOES SXR flux instead reflects the maximum intensity of thermal brehmsstrahlung emission from coronal plasma.  Since, in theory, the two classification systems (GOES class and impulsiveness) represent different physical mechanisms, we do not necessarily expect a relationship between the two.

\subsection{Discussion and limitations of impulsiveness classification}\label{sec:impclasslimit}

In this section, we discuss the limits of the use of the impulsiveness index for classification of solar flares.  In our development of the impulsiveness index, we are constrained to the timespan corresponding to the availability of MEGS-A data, between 2010 April 30 and 2014 May 26. We are also limited by the constraints of the \verb+RibbonDB+ catalog, which includes flares of GOES class C1.0 or higher within 45$^o$ of the central meridian \citep{kazachenko2017}.  We have developed the impulsiveness index indiscriminately for flares of GOES X-ray class C, M, or X, as shown in Figure \ref{imp_goes_recrate}(b), and have found no relationship between the impulsiveness index and peak GOES SXR flux.  We have not distinguished between eruptive and confined flares \citep[e.g.][]{kazachenko2023}.  Finally, we are limited to only consider flares with sufficient S/N in order to confidently calculate the impulsiveness index.

We developed our procedure for flare detection and vetting, in part, to avoid the restrictions on flare selection which, for example, \cite{fajardo-mendieta_2016} faced in their development of an ``impulsivity" parameter.  In their work, ``impulsivity" is defined as the inverse of the impulsive phase duration. \cite{fajardo-mendieta_2016} do not consider flares of GOES class C because of the possibility of an ill-defined peak intensity, and do not consider X class flares because of the possibility that a single recorded event may actually be a combination of many smaller flares. Regarding the exclusion of GOES class C, we are not similarly limited in our event selection, as the procedure for event selection only considers flares for which the start, peak, and end times of an event are clearly defined.  Regarding the exclusion of X-class flares, there is the possibility that our sample has included light curves which result from many smaller events rather than one large event.  We do not expect this concern to affect a significant number of events, particularly since our method excludes events which overlap so significantly that the flare parameters can not be determined.

We note that the selection of the line (or continuum) flux used to develop the impulsiveness index has significant impact on the behavior of the parameter.  We choose the SDO/EVE 304 \AA\ line \citep{milligan2012} for the reasons discussed in Section \ref{sec:data}: the line sensitivity to chromospheric emission, the availability of data, lack of saturation, and the relatively fast 10-second time cadence.  However, the 304 \AA\ line includes some contribution from loop-filling plasma \citep[e.g.][]{kuridze2013,delzanna2020} and could display longer decay phase times than other lines or continuum fluxes that capture chromospheric emission.  In Section \ref{highimpev}, we see that for one high impulsiveness event occurring on 4 April 2014, the 304 \AA\ light curve peaks after HXR emission has declined (and, therefore, the impulsive phase as traditionally defined has concluded). Regardless, because of the significant contribution from chromospheric ribbon emission, we consider the 304 \AA\ line as an appropriate initial choice for development of the impulsiveness index, although future work may reveal that there is a better option.

As described above, we have found little to no relationship between impulsiveness and other flare quantities, such as, for example, flare duration, rise phase duration, decay phase duration, and GOES SXR flux.  Impulsiveness index and peak reconnection flux have a moderate monotonically increasing relationship, and events with high peak reconnection rates disproportionately belong to the high impulsiveness category.  These results may be interpreted in two ways.  First, the impulsiveness index may capture a quantity that is not reflected in these other flare parameters.  By considering the intensity and duration of the impulsive phase in the EUV, the impulsiveness index may be an effective measure of the prominence of the impulsive phase, or the ``suddenness" of energy deposition in the chromosphere.  Based on the development of the impulsiveness index for stellar flares by \cite{kowalski2013}, we expect a connection between the impulsiveness of a light curve and the fundamentals of energy flux into the lower atmosphere, despite the complications and variability in magnetic configuration, ribbon motion, and flare development.  That events with high maximum reconnection rate are almost exclusively within the ``high-impulsiveness" category may indirectly be evidence that such a connection exists, since the energy deposited into the chromosphere is mostly released in the corona during reconnection.  

On the other hand, we must consider the possibility that classification based on impulsiveness is an oversimplification given the diversity of flaring regions and variability in the flaring process. This would introduce a randomness to the sample which could also result in the lack of correlation occurred.  In order to resolve this uncertainty, it would be useful to compare impulsiveness to other flare properties, particularly related to the white-light continuum radiation and the deposition of energy in the deeper chromosphere \citep[for example, spectral attributes such as those discussed by][]{kowalski2013}.

\section{Detailed analysis of selected events}\label{sec:casestudy}

In this section, we select six case study events of a range of impulsiveness values and analyze ribbon motion and flare light curves.

\subsection{Geometrical analysis of flare ribbons}

To analyze the flare ribbon motion for case study analysis, we use SDO/AIA and SDO/HMI observations.  We locate and track the evolution of the positive and negative polarity ribbons relative to the PIL.  From the relative motion of the ribbons, we derive a proxy for magnetic shear and estimate the strength of the reconnection component of the electric field.

\subsubsection{Determination of the polarity inversion line}

\begin{figure}
    \centering
    \includegraphics[width=.65\linewidth]{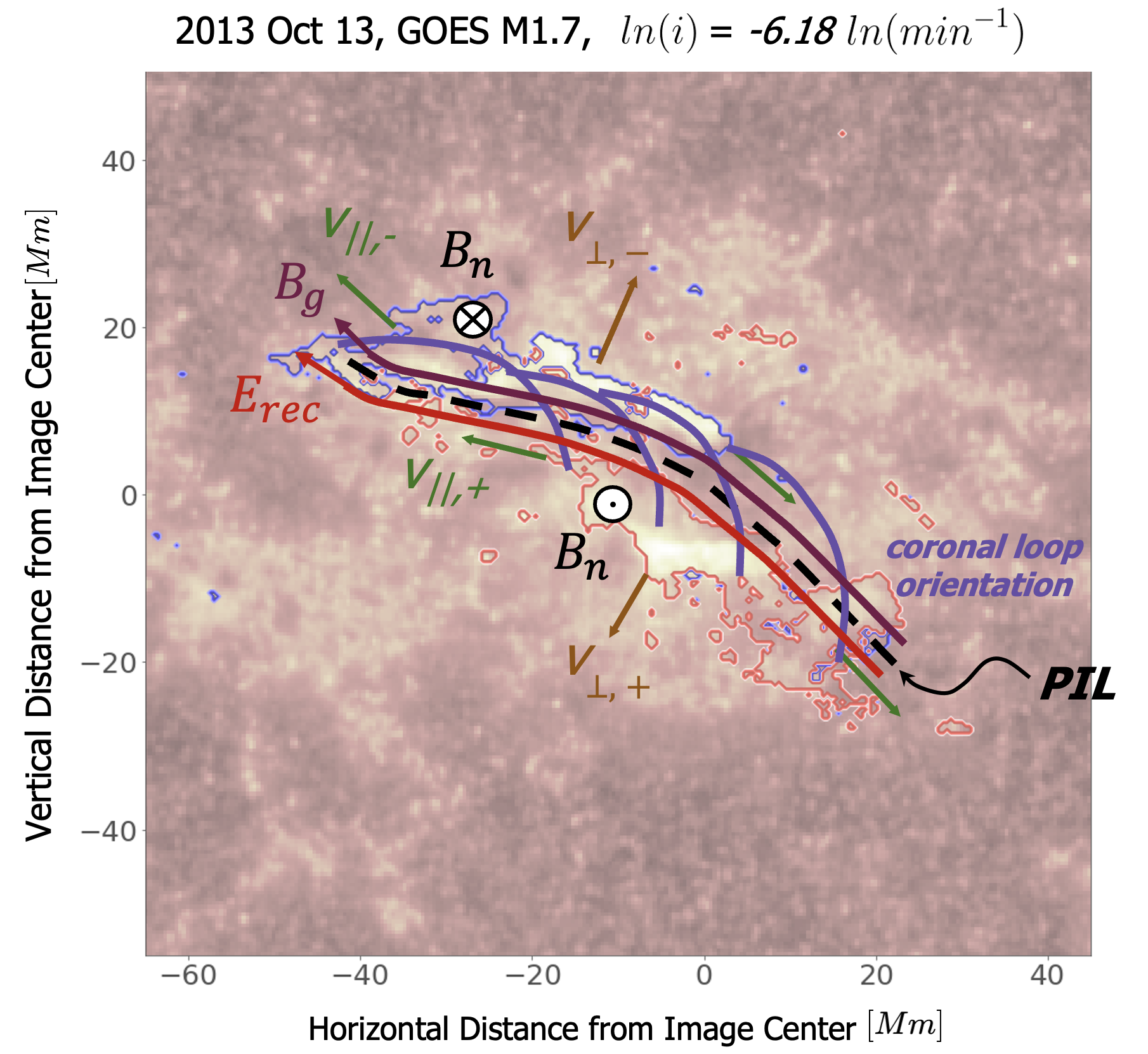}
    \caption{Flare ribbon magnetic properties overlaid onto SDO/AIA 1600 \AA\ image. Blue contours represent regions of negative polarity magnetic field.  Red contours represent regions of positive magnetic polarity.  $\mathbf{B}_n$ is the line-of-sight magnetic field.  The black dotted line indicates the location of the PIL, theoretically located directly under the reconnection electric field vector $\mathbf{E}_{rec}$ in the current sheet (red curved arrow).  In this case, $\mathbf{E}_{rec}$ is parallel to the guide field component of the magnetic field $\mathbf{B}_g$ (dark magenta curved arrow). Brown arrows show the PIL-perpendicular ribbon motion ($\mathbf{v}_{\perp,-},\mathbf{v}_{\perp,+}$), and green arrows the PIL-parallel motion ($\mathbf{v}_{\parallel,-},\mathbf{v}_{\parallel,+}$).  Light violet curves show the approximate sub-reconnection-region coronal loop orientation.  The angle between these loops and the overlying reconnection electric field demonstrates the significant amount of magnetic shear in this active region.  The background is an image from the SDO/AIA 1600 \AA\ channel.}
    \label{fig:examplegeo}
\end{figure}

We use the PIL as a geometrical reference for the parallel and perpendicular motion of chromospheric flare ribbons.  In the 3D solar flare model, the PIL lies under the coronal arcade of magnetic field loops, tracing the projection of the RCS on the solar surface.  As shown in Figure \ref{fig:examplegeo}, the reconnection electric field component $\mathbf{E}_{rec}$, the macroscopic electric field along the current sheet, is directed parallel to the PIL.  $\mathbf{E}_{rec}$ is sometimes used as a proxy for reconnection rate \citep[e.g.][]{qiu_2023}. We determine $\mathbf{E}_{rec}$ through the process described in Section \ref{ribcalc} and discussed in \cite{qiu2017}.  For 2.5D and 3D models of solar flares, the reconnection electric field is either parallel or antiparallel to the direction of the guide field component of the local magnetic field near the current sheet \citep{forbes2000,qiu2002,qiu2017,arnold2021,dahlin2021}.  

Inspired by \cite{wang2019}, we apply the semi-autonomous method described below for PIL identification, using the SDO/AIA 1600 \AA\ ribbon masks and SDO/HMI magnetograms:

\begin{enumerate}
    \item Derive the final cumulative mask in the series of images corresponding to the flare, with c = 8 as defined in \cite{kazachenko2017} and Equation \ref{eq:recrate}.
    \item Separate pixels in these images based on polarity (provided by SDO/HMI images) and place positive and negative ribbons into separate arrays.
    \item Remove spurious pixels not present in main ribbon areas, detected by visual inspection, with a variable criterion for whether a pixel is too far from the main ribbon area to be associated with the main regions of chromospheric energy deposition.
    \item Convolve the positive and negative ribbon pixels separately with a Gaussian function of flare-specific width.  This step requires oversight by the user, as ribbon geometry and active region size will inform the most appropriate Gaussian width.  Our method differs from that of \cite{wang2019} in that we do not universally use a Gaussian width of 10 pixels.  The width of the Gaussian which we convolve with the SDO/HMI images varies based on the spatial extent and proximity of the flare ribbons.  Ribbons that are narrower and closer together require a smaller Gaussian width in order to correctly identify the PIL.
    \item Multiply the convolved positive and negative ribbon masks; the non-zero region in the resulting image defines the PIL mask, a narrower curve-like shape between the ribbon masks.
    \item Fit a polynomial function (default $4^{th}$ order) to the PIL mask using Python's \texttt{numpy.polyfit} package.  The resulting curve of best fit is used as the PIL in further geometrical analysis.  We have added this step, which \cite{wang2019} do not perform, because our analysis requires an approximation of the location of the PIL with pixel-scale uncertainty in order to track small variations in ribbon motion.
\end{enumerate}

The position of the PIL, like the topology of the magnetic field, is assumed to be constant throughout the flare, as $B_z$ does not change appreciably during the period of emission in EUV wavelengths.  In Section \ref{ribcalc}, we describe the method by which we track ribbon pixels according to their position relative to the PIL by isolating perpendicular and parallel ribbon motion.

\subsubsection{Calculating ribbon motion and the reconnection electric field}\label{ribcalc}

We decompose ribbon motion into PIL-perpendicular and PIL-parallel components.  Previous studies have done a similar decomposition \citep[e.g.][]{qiu2010,cheng2012,qiu2017,qiu2022}.  For each 24-second time step in the SDO/AIA 1600 \AA\ image data, we establish the PIL-parallel and PIL-perpendicular distance.  The time evolution of these quantities in each ribbon represents the geometrical evolution of the flare in this wavelength.  The quantities are discussed in Section \ref{sec:caseresults} for a selection of six flares.

We analyze PIL-perpendicular motion (``separation", as in the brown arrows of Figure \ref{fig:examplegeo}) using the instantaneous ribbon masks.  For each image, we isolate the positive and negative masks and remove spurious pixels (those which are separate from the main ribbon areas). Then, for each ribbon pixel, we determine the distance to the closest point on the PIL-representative polynomial.  The reported PIL-perpendicular ribbon distance ($l_{\perp,\pm}$) in each polarity for each time step is the median of all minimum pixel-PIL distances.  Studies have traditionally used the mean pixel-PIL distance as a proxy for PIL-perpendicular ribbon motion.  We find that, for the case study flares analyzed below, use of the median pixel-PIL distance is more representative of ribbon distance.

We determine PIL-parallel motion (``elongation", as in the green arrows of Figure \ref{fig:examplegeo}) using the cumulative ribbon masks.  Similar to our scheme to determine the extent of PIL-perpendicular motion, we isolate flare ribbons for each time step and remove spurious pixels according to a variable criterion, based on visual detection, for inclusion in the ribbon.  We refer to ``east" and ``west" ends of a ribbon as ``left" and ``right," respectively.  The method automatically locates the extreme limits (left and right) of each ribbon, in each polarity, for each time step.  We project the left and right coordinates onto the PIL, and record the curve length along the PIL between the left and right corresponding points on the line representative of the PIL to be the PIL-parallel extent of each ribbon for that time step ($l_{\parallel,\pm}$).  

Here, we justify our use of the instantaneous ribbon masks for PIL-perpendicular motion and the cumulative ribbon masks for PIL-parallel motion.  Compared to the instantaneous ribbon area, the cumulative ribbon area more effectively captures the PIL-parallel extent of chromospheric emission during the flare.  Use of the cumulative ribbon masks for determination of PIL-parallel ribbon motion produces lengths more in agreement with the observed magnetic structure.  The cumulative masks identify the orientation of newly excited coronal loops as well as those still present in the active region but decaying in strength.  Our use of instantaneous masks in identifying the magnitude of PIL-perpendicular ribbon motion essentially locates the outer edge of ribbon separation, which we find more effectively captures the PIL-perpendicular extent of coronal loops.  Studies which decompose ribbon motion have not typically used the cumulative and instantaneous ribbon masks for different use cases.  For example, \cite{qiu2010} determine both PIL-parallel extent and PIL-perpendicular distance using the newly brightened ribbon front.

Having established the PIL-perpendicular component of the ribbon motion in each polarity, we take the time derivative of the PIL-perpendicular ribbon motion to quantify the ribbon velocity perpendicular to the PIL.  Using the ribbon velocity, we can determine the electric field strength along the RCS according to the method of \cite{qiu2017}.  We approximate the reconnection electric field $\mathbf{E}_{rec}$ using the equation for the Lorentz force:

\begin{equation}\label{recE}
\mathbf{E}_{rec} = -\mathbf{v}_{in}\mathbf{\times} \mathbf{B}_{in} \approx \mathbf{v}_\perp \mathbf{\times} \mathbf{B}_n,
\end{equation}

\noindent where $\mathbf{v}_{in}$ and $\mathbf{B}_{in}$ are the plasma inflow velocity and magnetic field vector at the edge of the small-scale diffusion region, $\mathbf{v}_{\perp}$ is the PIL-perpendicular ribbon velocity, and $\mathbf{B}_n$ is the strength of the normal component of the magnetic field measured by SDO/HMI (as in Figure \ref{examplecurve}).  We determine the direction of the reconnection electric field using Equation \ref{recE}, which assumes (1) that the inflow velocity in the RCS and the rate at which new ribbons appear in chromospheric lines are equal in magnitude and opposite in direction, and (2) that the line-of-sight magnetic field, $\mathbf{B}_n$, and the magnetic field vector into the RCS, $\mathbf{B}_{in}$, are of the same magnitude. 
Since $\mathbf{v}_{\perp}$ and $\mathbf{B}_{n}$ are orthogonal, we use the scalar equation $E_{rec} \approx v_{\perp}B_n$ to estimate the strength of the reconnection component of the electric field \citep{qiu2002,forbes2000}.  We note that $v_{\parallel}$ as defined in \cite{qiu2002} and \cite{forbes2000} is analogous to $v_{\perp}$ here.  We believe that the notation we use, which agrees with that of \cite{qiu2017}, is more appropriate for our analysis in order to properly distinguish between PIL-parallel and PIL-perpendicular motion.

\subsubsection{Determination of magnetic shear}

The presence of a magnetic field with a large amount of magnetic shear suggests that the guide field component of the coronal magnetic field is strong. The guide field may play a role in suppression of particle precipitation, thereby having an effect on the timing of energy release (and, possibly, the impulsiveness) of a flare. 
 
 We use a method similar to that of \cite{qiu2017} in our determination of magnetic shear, and also draw inspiration from the theoretical studies of \cite{arnold2021}  and \cite{dahlin2021}. When projected onto the solar surface, the coronal arcade of magnetic field loops in a highly sheared configuration will lie more parallel to the PIL.  As a result, the conjugate footpoints of these loops will be offset along the PIL.  Since the UV footpoints represent the regions of chromospheric condensation from particles propagating along the arcade, the angle of the line relative to the PIL connecting two pixels at the positive and negative footpoints of the same coronal loop is a good representation of the amount of magnetic shear in the active region (see Figure \ref{fig:examplegeo}).  We can therefore track the relative positions of corresponding negative and positive polarity footpoints in the flare ribbons to estimate magnetic shear.

An efficient way to correctly pair the spatial regions representing the two feet of the same coronal loop is to identify pixels at both ends of the instantaneous ribbon images in each polarity.  The new pixels in each frame at these locations are most likely associated with the same coronal loop.

Here we outline the details of our determination of shear.  We use the instantaneous ribbon masks in each frame, remove spurious, localized points included in the masks, and locate the extreme ends of each ribbon in each polarity.  We then calculate the PIL-perpendicular and PIL-parallel components of the distance between these points.  The ratio of the PIL-parallel distance to the PIL-perpendicular distance is a proxy for the guide field ratio (GFR), a measure of magnetic shear similar to that presented in \cite{dahlin2021}.  The GFR as determined by the position of flare ribbon fronts is proportional to the strength of the guide field component of the magnetic field relative to the reconnecting component of the magnetic field \cite[e.g.][]{zhou_2009,qiu_2023}.  It is important to note that our use of the term ``guide-field ratio" here is not a direct measure of magnetic field strengths, but rather a proxy for these strengths using ribbon position.  

A larger GFR corresponds to a higher shear.  We define, in Equation \ref{gfrangle}, the shear angle $\theta$, the angle between the coronal loops and a line perpendicular to the PIL, relative to the GFR:

\begin{equation}\label{gfrangle}
\theta = 90^o - tan^{-1}\left(\frac{l_{\perp,+}+l_{\perp,-}}{l_{\parallel,+}+l_{\parallel,-}}\right) = 90^o - tan^{-1}\left(GFR^{-1}\right),
\end{equation}

\noindent where $\theta$ is the shear angle,  $l_{\perp,+}$ ($l_{\perp,-}$) is the perpendicular distance from the PIL to the end of the positive (negative) ribbon, and $l_{\parallel,+}$ ($l_{\parallel,-}$) is the PIL-parallel distance from the end of the positive (negative) ribbon to the point on the PIL which intersects with a line connecting one end of the negative ribbon to the analogous end of the positive ribbon.  A large (small) $\theta$ corresponds to a high (low) amount of magnetic shear.

Equation \ref{gfrangle} is applied to both ends of the ribbons separately. Our method returns values for the left and right shear angles of each image frame.  We average the left and right shear angles to estimate the magnetic shear present at a given point in time. 

\subsection{Selection of events for case study analysis}

We select events for case study analysis from the population of all flares in the \verb+RibbonDB+ catalog with discernible light curves in the 304 \AA\ line.  Table \ref{tab:evsel} details the flare events chosen.  We selected the six events listed in Table \ref{tab:evsel} according to the following criteria: (1) lack of saturation effects in the 1600 \AA\ full-disk images; (2) clear linear or curvilinear two-ribbon structure in the 1600 \AA\ full-disk images; (3) easily discernible PIL-relative ribbon motion; (4) an impulsiveness index belonging to one of three categories, ``low" ($\ln[i] <-6 \ln{[min^{-1}]}$), ``mid" ($-6 < \ln[i] < -5 \ln{[min^{-1}]}$), or ``high" ($\ln[i] > -5 \ln{[min^{-1}]}$), with the final sample of six flares including at least two low-impulsiveness and two high-impulsiveness events, in order to compare flares belonging to the most extreme impulsiveness categories. Criteria (1)-(3) are necessary in order to quantify ribbon motion and infer coronal loop orientation. We vetted the sample of flares provided by \verb+RibbonDB+ extensively in search of the events that best satisfy these criteria, first surveying the sample of 480 best-performing flares and only considering flares outside of this subsample if the criteria were not met for a particular impulsiveness category. The result is our final sample of flares.  Events Mid1 and High2 do not belong to the sample of best-performing 480 flares as determined by the flare selection algorithm. Because of issues with saturation in the 1600 \AA\ images associated with higher-impulsiveness events, we selected these two events from the sample of 1368 flares and verified by inspection that they satisfy the criteria above for analysis of spatial evolution.

We also note whether the chosen events are classified as eruptive (associated with a coronal mass ejection, CME) or confined (with no opening of field lines and therefore no observed coronal mass ejection) by consulting the SOHO LASCO CME catalog.\footnote{The LASCO CME catalog is generated and maintained at the CDAW Data Center by NASA and The Catholic University of America in cooperation with the Naval Research Laboratory. SOHO is a project of international cooperation between ESA and NASA.}  As in \cite{li_2019}, if a CME occurs within 2 hours of the listed start time of the flare, and originates in the same quadrant, we identify that flare as ``eruptive."  Although many two-ribbon flares are eruptive, this is not universally the case \citep{masson_2009,kazachenko2023}. Events Low1, Low2, and High1 are eruptive; events Low3, Mid1, and High2 are confined.

By selecting flares in part based on clear curvilinear ribbon structure, we have limited our analysis to six flares with the classical two-ribbon structure. Note, however, that here are many ribbon configurations associated with solar flares \citep{masson_2009,dalmasse_2015,joshi_2019}.  Our objective is to analyze the ribbon geometry and flare energy release patterns for flares of the kind produced by the simulations of \cite{dahlin2021} in order to study in detail, for a small selection of flares, the connection between magnetic configuration and energy release predicted by these simulations, for a range of impulsiveness values.  We have therefore restricted ourselves to selection of two-ribbon flares.

\begin{deluxetable}{|c|c|c|c|c|c|c|c|}
\tabletypesize{\footnotesize}
\tablewidth{0pt}

 \tablecaption{ Final selection of flares for case study analysis \label{tab:evsel}}
 \tablehead{\colhead{Flare ID} &
 \colhead{Date} & \colhead{Time (UT)} & \colhead{GOES Class} & \colhead{$\ln(i)$ [$\ln{(min^{-1})}$]} & \colhead{CME} & \colhead{Figure} & \colhead{Section}}
 
\startdata
 Low1 & 2013 Oct 13 & 00:12 & M1.7 & -6.18 (Low) & Yes & \ref{fig:low1low2}(a)-(c) & \ref{lowimpev} \\
    \hline
 Low2 & 2013 Oct 15 & 08:26 & M1.8 & -6.34 (Low) & Yes & \ref{fig:low1low2}(d)-(f)  & \ref{lowimpev} \\
    \hline
 Low3 & 2012 Nov 20 & 19:21 & M1.6 & -6.98 (Low) & No & \ref{fig:low3mid1}(a)-(c)  & \ref{lowimpev} \\
    \hline
 Mid1 & 2013 May 16 & 21:36 & M1.3 & -5.06 (Mid) & No & \ref{fig:low3mid1}(d)-(f)  & \ref{midimpev} \\
    \hline
 High1 & 2014 Apr 4 & 03:43 & C3.6 & -4.73 (High) & Yes & \ref{fig:high1high2}(a)-(c) & \ref{highimpev}   \\
    \hline
 High2 & 2014 Apr 15 & 08:53 & C1.3 & -4.18 (High) & No & \ref{fig:high1high2}(d)-(f) & \ref{highimpev}  \\
\hline
\enddata
 \vspace{-0.5cm}
 \tablecomments{sample comment}
\end{deluxetable}

\subsection{Results of case study analysis}\label{sec:caseresults}

\begin{figure}
    \centering
    \includegraphics[width=.8\textwidth]{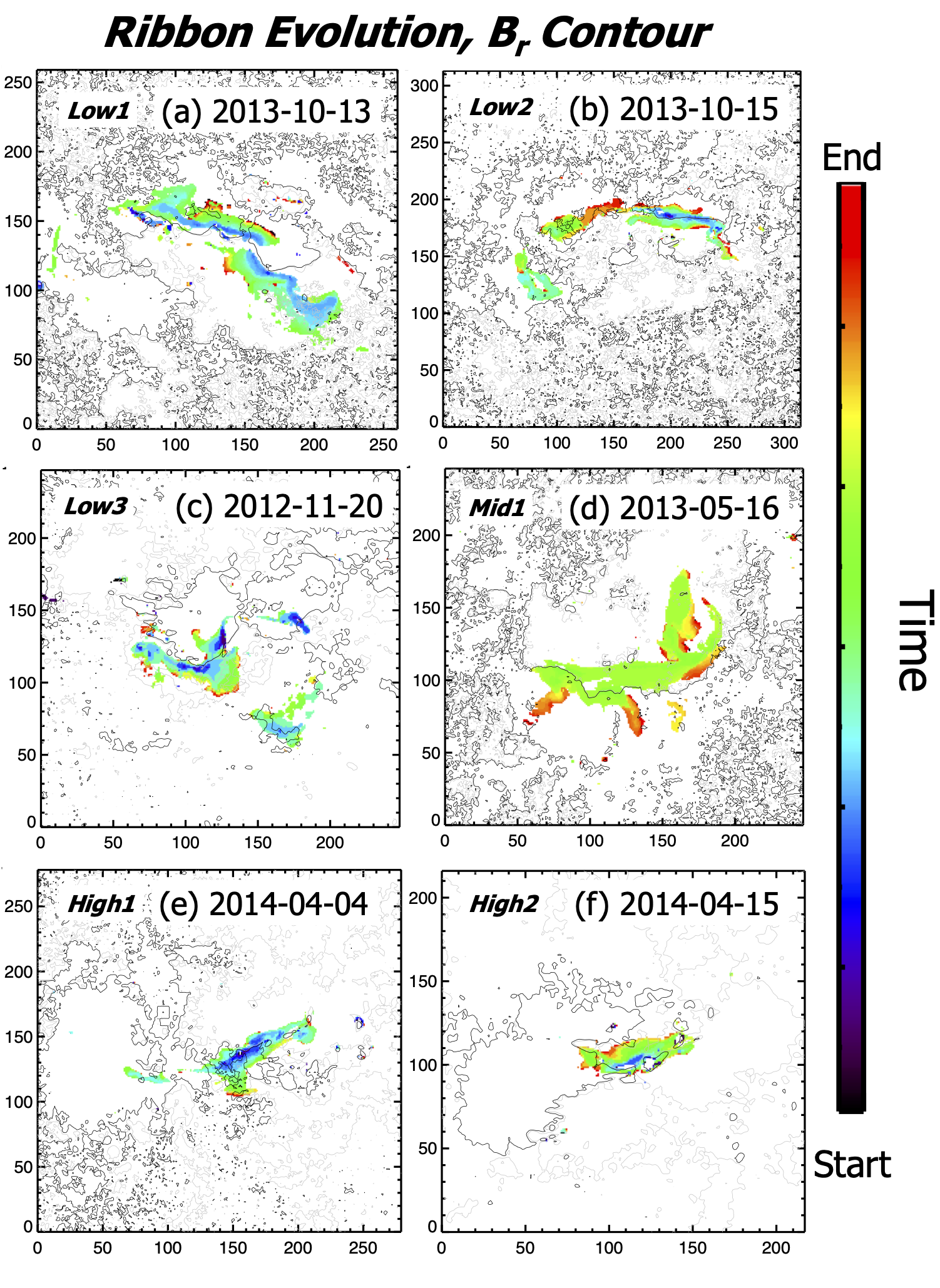}
    \caption{Ribbon evolution and radial magnetic field contours for the six case study events.  Colors indicate the percentage of total flare time which has passed before a pixel shows emission above the masking threshold in the 1600 \AA\ line.  Contours reflect the magnitude of the radial magnetic field.  Panels (a), (b), and (c) correspond to the low-impulsiveness events; panel (d) to the mid-impulsiveness event; and panels (e) and (f) to the high-impulsiveness events.  Axis label units are in pixels.}
    \label{fig:ribbonev}  

\end{figure}

In this section we compare the ribbon geometry and evolution, reconnection electric field strength, and EUV/HXR light curves for the events listed in Table \ref{tab:evsel}, corresponding to low (3 flares), mid (1 flare), and high (2 flares) impulsiveness  categories. Figure \ref{fig:ribbonev} shows the time evolution of ribbon masks for six case study events of varying impulsiveness. 

\subsubsection{Low impulsiveness events: 2013 Oct 13 (GOES M1.7), 2013 Oct 15 (GOES M1.8), 2012 Nov 20 (GOES M1.6)}\label{lowimpev}

\begin{figure}
    \centering
    \includegraphics[width=.9\linewidth]{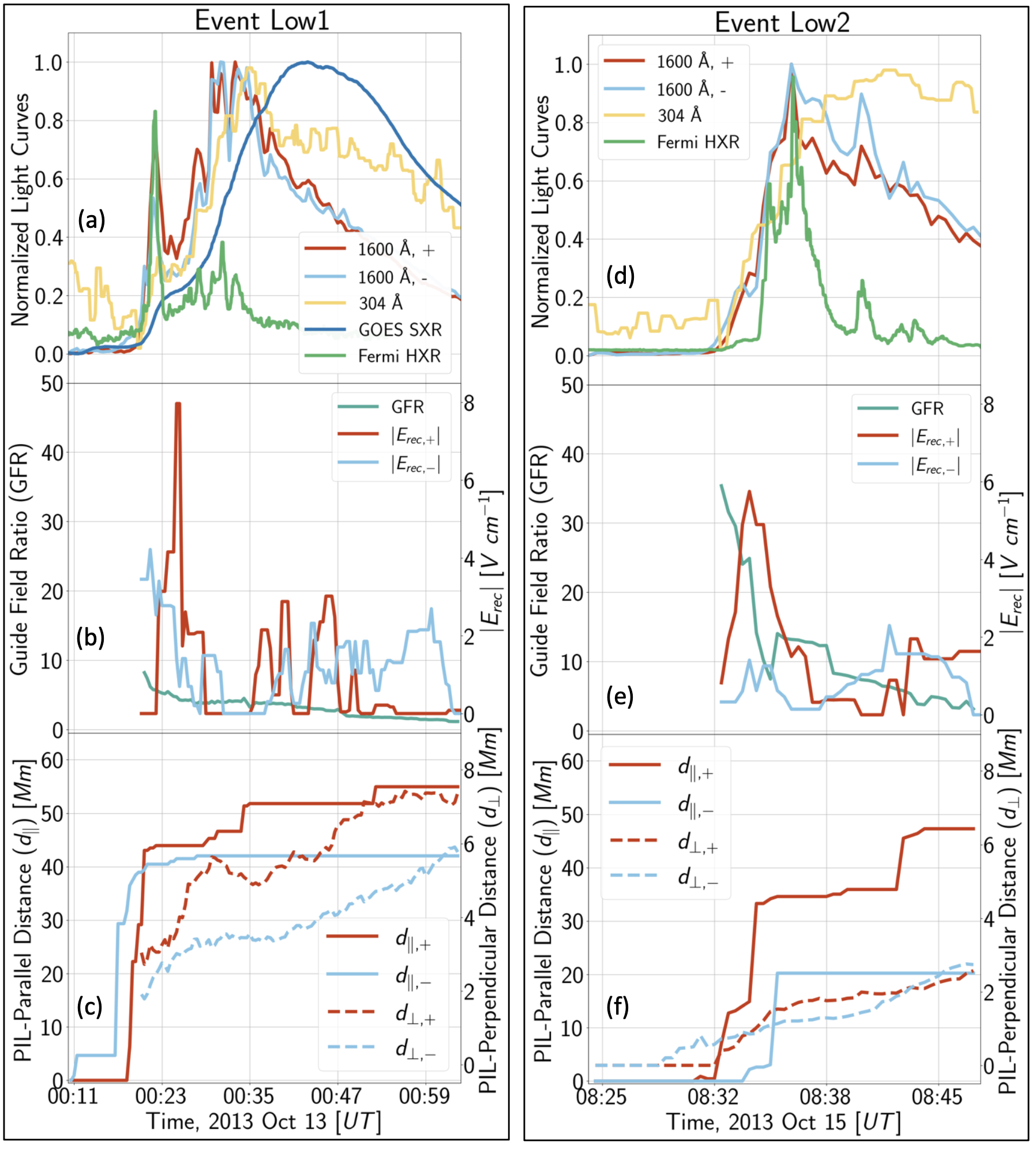}
    \caption{Details of flare evolution for events Low1 and Low2.  (a)-(c) Low impulsiveness ($\ln[i] = $ -6.18 $\ln[{min^{-1}}]$) GOES Class M1.7 flare on 2013 October 13 (flare Low1). (d)-(f) Low impulsiveness ($\ln[i] =$ -6.34 $\ln[{min^{-1}}]$) GOES class M1.8 flare occurring on 2013 October 15 (flare Low2). In panels (a) and (d), the green line corresponds to the background-subtracted HXR counts from 25 to 300 keV; the red and light blue lines correspond to the 1600 \AA\ integrated counts in positive and negative polarity regions, respectively; the yellow line corresponds to the 304  \AA\ Sun-as-a-star flux; and the dark blue line corresponds to the GOES XRS 1-8 \AA\ SXR flux.  In panels (b) and (e), the red and blue lines correspond to the reconnection electric field strength determined from the positive and negative ribbons, respectively; the green line represents the guide field ratio (GFR), a proxy for magnetic shear. In panels (c) and (f), ribbon motion relative to the polarity inversion line (PIL) is displayed; the solid (dashed) red and blue curves correspond to PIL-parallel (PIL-perpendicular) ribbon motion in positive and negative polarity regions, respectively.}
    \label{fig:low1low2}  

\end{figure}

Panels (a)-(c) of Figure \ref{fig:low1low2} show the development of the 2013 October 13 GOES class M1.7 flare of low impulsiveness index $\ln(i) = $ -6.18 $\ln({min^{-1}})$. The HXR and 1600 \AA\ emission are co-temporal, and have a sharp peak early in the rise phase, at about 00:22 UT.  The rise phase continues until around 00:33 UT, when the HXR and 1600 \AA\ light curves peak, roughly 3 minutes before the 304 \AA\ light curve does.  Elongation of ribbons is limited to the early phase of the 1600 \AA\ emission, ending at roughly 00:20 UT, though the positive ribbon elongates for a slightly longer period.  Separation of ribbons persists through the decay phase.  However, in both ribbons, there is a brief cessation of perpendicular motion around 00:35 UT.  The ribbons separate before and after this time.  There is a decrease in the magnetic shear just before the peak in HXR emission.  The GFR continues to decrease slowly through the remainder of the rise phase and the entirety of the decay phase.  Although the rate of shear decrease is more significant early in the development of the flare, the quarter-max decay time, the time it takes for shear to decrease from its maximum value to 25\% of its maximum value relative to the minimum amount of shear, is 24.4 minutes.  This is a long period relative to the shear decay of other flares, suggesting that the guide field persists longer for this event.  The reconnection electric field is strongest during the rise phase and decreases after the peak EUV and HXR intensity. The ribbon evolution of this event is shown in Figure \ref{fig:ribbonev}(a). The ribbons are geometrically linear, and the ``spreading" of ribbons along and away from the PIL is evident.  Earlier ribbons (blue) are offset and close to the PIL, while later ribbons are not offset, and are farther from the PIL.  Coronal loop structure in, for example, the 171 \AA\ line (not pictured), yields the same results:  a classical two-ribbon structure, with new coronal loops appearing first along the PIL and then expanding away from the PIL, producing the strong-to-weak shear pattern observed.

Panels (d)-(f) of Figure \ref{fig:low1low2} show results for the second low-impulsiveness event (Low2), occurring on 2013 October 15, of GOES class M1.8. For this flare, $\ln(i) = $  -6.34 $\ln({min^{-1}})$. 1600 \AA\ and HXR emission peak co-temporally, while the 304 \AA\ curve has a peak delayed by about 6 minutes and has a much longer duration rise phase (roughly twice the rise time of the 1600 \AA\ curve).  As in the case of the previous flare, PIL-parallel ribbon motion is limited to the rise phase of the EUV and HXR light curves, while the PIL-perpendicular ribbon motion occurs at a constant rate through both the rise and decay phases.  The amount of magnetic shear in the active region decreases substantially just before a gradual HXR light curve rise.  The quarter-max decay time of magnetic shear is 2.4 minutes, significantly shorter than that of event Low1. However, we note that the absolute magnitude of shear remains high ($GFR \approx 10$) in this flare despite the initial fast decrease in shear. The reconnection electric field determined from the positive ribbon is strongest during the rise phase.  The reconnection electric field determined from the negative ribbon is relatively constant throughout the flare.  The ribbon evolution of this event is shown in Figure \ref{fig:ribbonev}(b). Once again, most of the ribbon area lies parallel to the PIL. The flare ribbons clearly separate from the PIL later in the flare.

\begin{figure}
    \centering
    \includegraphics[width=.9\linewidth]{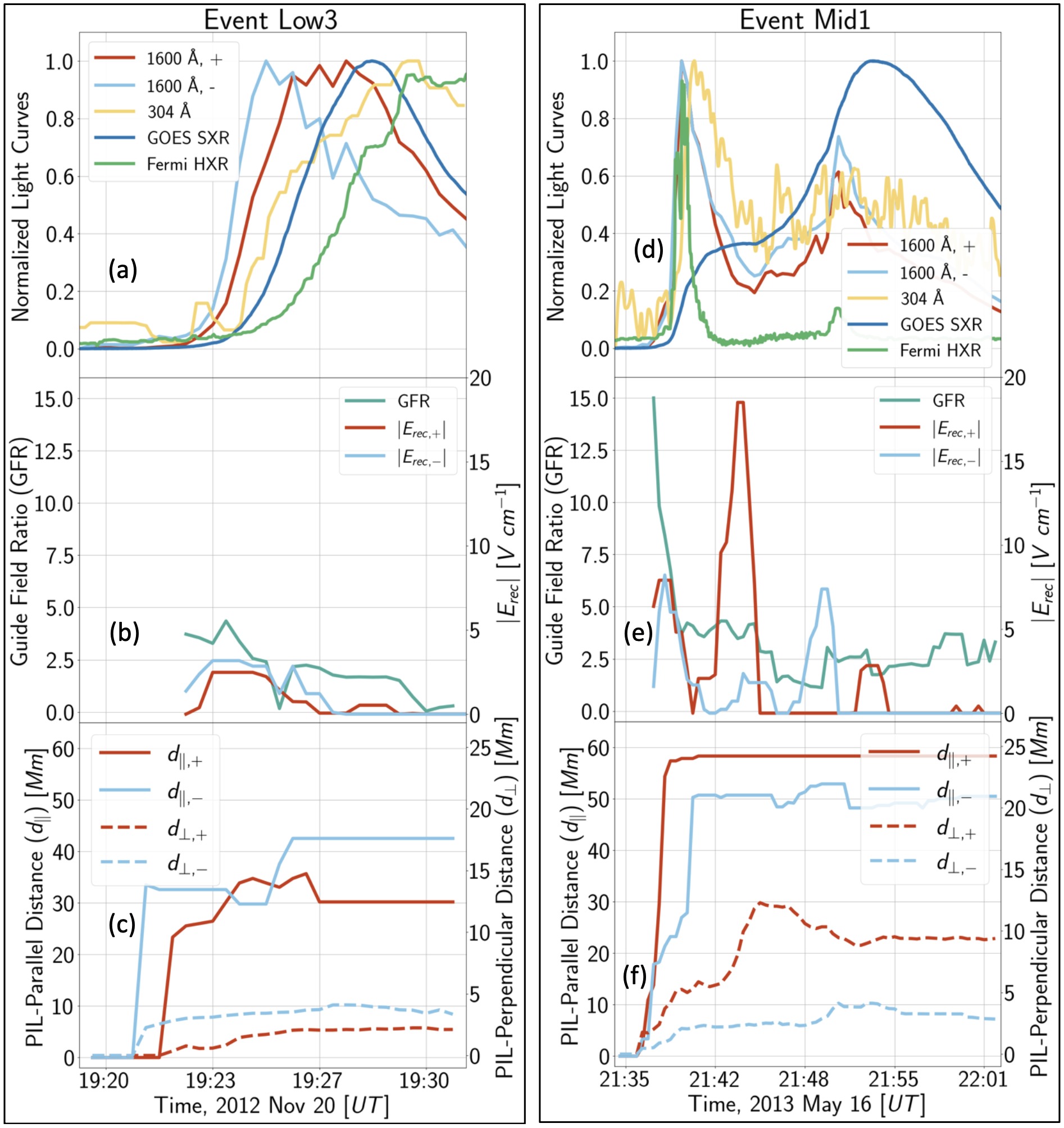}
    \caption{Details of flare evolution for events Low3 and Mid1. (a)-(c) Low impulsiveness ($\ln[i] = $-6.98 $\ln[{min^{-1}}]$) GOES class M1.6 flare occurring on 2012 November 20 (flare Low3). (d)-(f) Mid-impulsiveness ($\ln[i] =$-5.06 $\ln[{min^{-1}}]$) GOES class M1.3 flare occurring on 2013 May 16 (flare Mid1). The figure is organized as described in the caption of Figure \ref{fig:low1low2}.}
    \label{fig:low3mid1}  

\end{figure}

Panels (a)-(c) of Figure \ref{fig:low3mid1} show results for the low-impulsiveness 2012 November 20 GOES class M1.6 flare (Low3) of low impulsiveness $\ln(i) = $ -6.98 $\ln({min^{-1}})$.  Low3 has a slightly different pattern compared to the other two low-impulsiveness case studies in that the 304 \AA\ and HXR emission are cotemporal, with the 1600 \AA\ integrated light curve evolving roughly 3 - 5 minutes before.  Ribbon length increases quickly in the very early stages of flare development, and continues to increase slightly until the peak in EUV light curves.  PIL-perpendicular ribbon motion and shear are relatively constant throughout, although there is a sharper increase in the PIL-perpendicular motion of the negative polarity ribbon during the early rise phase.  The GFR is low ($\approx 1 - 5$), particularly when compared to events Low2 or Mid1; however, the quarter-max decay time (6.4 minutes) is longer than those of events Low2 and Mid1 but shorter than that of event Low1.  Notably, there is no sharp decrease in shear; rather, the shear decay occurs more slowly.  This, as discussed in Section \ref{sheardisc}, may contribute to the gradual increase in HXR emission when compared to most of the other case studies here.  The reconnection electric field is strongest during the rise phase, and decreases after the peak in EUV and HXR emission.  
The ribbon evolution of this event is shown in Figure \ref{fig:ribbonev}(c). 

\subsubsection{Mid impulsiveness event: 2013 May 16 (GOES M1.3)}\label{midimpev}
Panels (d)-(f) of Figure \ref{fig:low3mid1} show results for the 2013 May 16 GOES class M1.3 flare, of mid-impulsiveness $\ln(i) = $ -5.06 $\ln({min^{-1}})$.  The 1600 \AA\ emission and HXR flux peak simultaneously at around 21:40 UT, and show some late-phase flaring. The late-phase flaring is less apparent in the 304 \AA\ 
light curve, which peaks later than the other two wavelengths (at roughly 21:41 UT).  As in events Low1 and Low2, the period of PIL-parallel ribbon motion is limited to the rise phase of the EUV and HXR emission.  PIL-perpendicular ribbon motion extends throughout, but in this case occurs mostly during the rise phase.  The sharp increase in PIL-perpendicular positive-polarity ribbon motion during the decay phase of the flare is due to the emergence of a large positive-polarity flaring region perpendicular to the PIL.  The structure is included in our analysis because of its spatial magnitude relative to the active region itself.  However, the large PIL-perpendicular flare region is likely the result of a more complicated magnetic structure than is typically presented in a 2.5D or 3D model, and may cause a substantial overestimation in the calculated PIL-perpendicular ribbon velocity and, in turn, the reconnection electric field.  

The quarter-max decay time for this flare is only 1.6 minutes, lower than that of all low-impulsiveness events.  The first significant rise in HXR emission occurs after most of the shear in the active region has subsided.  There is a second, smaller reduction in shear around 21:46 UT, which occurs roughly 4 minutes before a second peak in HXR and EUV emission.  In a clear demonstration of the Neupert effect \citep{neupert_1968}, both HXR peaks (occurring at roughly 21:40 UT and 21:50 UT) occur during a rise in SXR emission.  The reconnection electric field is strongest during the rise phase, though there is a peak in the reconnection electric field determined from the negative ribbon during the decay phase.  The positive-polarity region of increased emission mentioned above also affects the calculated strength of the reconnection electric field determined from the positive ribbon. 
The ribbon evolution of this event is shown in Figure \ref{fig:ribbonev}(d).  The PIL is less discernible for this flare, but we determine that it is an almost horizontal line bisecting the lower arm of the ribbons, separating positive and negative polarities.

\subsubsection{High impulsiveness events: 2014 April 4 (GOES C3.6), 2014 April 15 (GOES C1.3)}\label{highimpev}

\begin{figure}
    \centering
    \includegraphics[width=.9\linewidth]{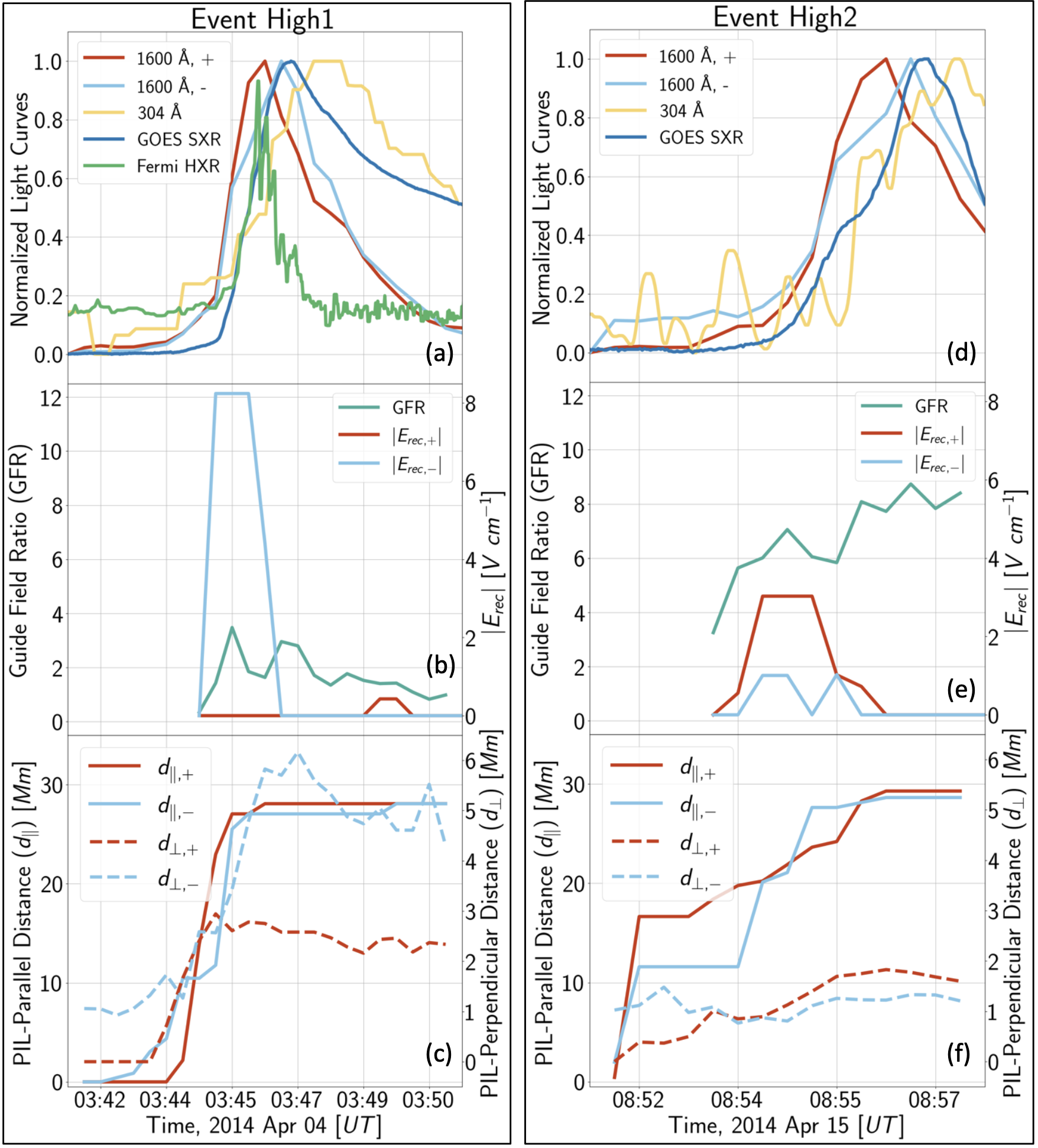}
    \caption{Details of flare evolution for events High1 and High2. (a)-(c) High impulsiveness ($\ln[i] = $-4.73 $\ln[{min^{-1}}]$) GOES class C3.6 flare occurring on 2014 April 4 (flare High1). (d)-(f) High impulsiveness ($\ln[i] = $ -4.18 $\ln[{min^{-1}}]$) GOES class C1.3 flare occurring on 2014 April 15 (flare High2). The figure is organized as described in the caption of Figure \ref{fig:high1high2}.}
    \label{fig:high1high2}  

\end{figure}

Panels (a)-(c) of Figure \ref{fig:high1high2} show the results of our analysis for the 2014 April 4 GOES class C1.3 flare with a high impulsiveness of $\ln(i) = $ -4.73 $\ln({min^{-1}})$.  The peak intensity of this event in the HXR and EUV lines is lower than in the low or mid-impulsiveness events.  However, the duration is also significantly shorter, producing the observed high impulsiveness value.  The duration of this event ($\Delta t \approx 5$ $min$) is closer to the temporal resolution of observations than the durations of the mid- and low-impulsiveness events ($\Delta t \geq 15$ $min$), making it difficult to resolve geometrical and energy release patterns.  The HXR and 1600 \AA\ light curves are roughly co-temporal, with a peak just before 03:46 UT.  The 304 \AA\ curve peaks a minute later, at about 03:48 UT.   Unlike the low- and mid-impulsiveness events, both components of PIL-relative motion (rather than just the PIL-parallel motion) are roughly complete at the end of the EUV rise phase.  The shear present during the event decreases slightly, concurrently with the increase in EUV and HXR emission, although the pattern is less clear than in other case studies due to the short timespan of the event.  Shear values are consistently low ($GFR \approx 1-4$). The HXR emission clearly shows a rise and decay pattern, though the curve does not rise as sharply as in events with a more significant decrease in shear during the EUV rise phase.  The ribbon evolution of this event is shown in Figure \ref{fig:ribbonev}(e).  

Panels (d)-(f) of Figure \ref{fig:high1high2} show results for the 2014 April 15 GOES class C1.3 flare of high impulsiveness $\ln(i) = $ -4.18 $\ln({min^{-1}})$.  The high impulsiveness of the event is likely due to the fast rise and decay of the flare, as the peak EUV emission is low.  Both PIL-perpendicular and PIL-parallel ribbon motion occur during the relatively long rise phase of the flare, though the negative ribbon shows little PIL-perpendicular motion. There is no discernible HXR emission above the background flux in the 25 - 300 keV band according to Fermi/GBM during the time period studied here.  Magnetic shear remains high and constant throughout the event ($GFR \approx 4-8$).  In fact, the coronal loops are consistently highly sheared throughout. Coronal loop structure in the 171 \AA\ line, assessed using the JHelioviewer software \citep[][not pictured]{muller2017} suggests a more complicated magnetic field topology when compared to the low and mid-impulsiveness events, with a less clear arcade structure and, potentially, a persistent coronal loop nearly parallel to the PIL, although it is difficult to determine the precise orientation of these highly sheared loops.  It is possible that the consistent and high amount of shear in this region suppresses particle acceleration (and therefore its proxy, HXR emission) as in the 3D MHD simulations of \cite{dahlin2021}. The ribbon evolution of this event is shown in panel (f) of Figure \ref{fig:ribbonev}.

\subsection{Discussion of case study analysis}

In this section, we discuss the results of our case study analysis in the context of recent work. 

\subsubsection{Motion of ribbons relative to PIL}
For events Low1, Low2, Low3, and Mid1, the PIL-parallel motion and PIL-perpendicular motion of the chromospheric flare ribbons follow the strong-to-weak shear pattern reported in past observations and simulations \citep{qiu2017,dahlin2021,qiu2022}.  We display the shear evolution in panels (b) and (e) of Figures \ref{fig:low1low2}, \ref{fig:low3mid1}, and \ref{fig:high1high2}, and the PIL-parallel and PIL-perpendicular distances in panels (c) and (f) of Figures \ref{fig:low1low2}, \ref{fig:low3mid1}, and \ref{fig:high1high2}. In events Low1 and Low2, the PIL-parallel ribbon motion occurs almost entirely just before the peaks of the 1600 \AA, 304 \AA, and HXR light curves, while PIL-perpendicular ribbon motion persists through the decay phase. Event Mid1, in Figure \ref{fig:low3mid1}(d)-(f), shows a similar pattern, as does event Low3, in Figure \ref{fig:low3mid1}(a)-(c), to a lesser extent.  In event Low3, the PIL-perpendicular motion of the negative ribbon is most significant during the rise phase.  The positive ribbon experiences steady PIL-perpendicular motion.  Most of the PIL-parallel ribbon motion occurs in the rise phase.  

The high-impulsiveness events depart slightly from the pattern of early PIL-parallel motion and late PIL-perpendicular motion of ribbons.  This may be due to their short duration or the complexity of the coronal loops associated with these events.  During event High1, as shown in Figure \ref{fig:high1high2}(a)-(c), both the PIL-parallel and PIL-perpendicular ribbon motion occur entirely in the rise phase of the flare.  During event High2, as shown in Figure \ref{fig:high1high2}(d)-(f), PIL-parallel motion ceases before the end of the rise phase, and neither ribbon experiences much PIL-perpendicular motion.  

Our observations of ribbon motion relative to the PIL are consistent with our calculated values of magnetic shear. In events Low1, Low2, Low3, and Mid1, the pattern is as follows: PIL-perpendicular motion occurs during both the rise and decay flare phases, increasing the PIL-perpendicular component of overlying loop length. PIL-parallel motion, on the other hand, occurs mostly during the rise phase.  During the flare decay phase, the PIL-parallel component of loop length is relatively constant. As a result of the simultaneous PIL-parallel and PIL-perpendicular ribbon motion, overlying magnetic field loops become more perpendicular to the PIL.  When the pattern of early elongation with early and late separation of ribbons is observed, we expect a decrease in measured magnetic shear which is more significant during the flare rise phase, as is the case in events Low1, Low2, Mid1, and to a lesser extent, Low3. 

The dynamics are more complicated in events High1 and High2.  In event High1, both PIL-parallel and PIL-perpendicular ribbon motion occur almost entirely during the rise phase.  In event High2, both types of ribbon motion occur throughout the flare.  As a result, the change in magnetic shear during these events is not as clear (see Section \ref{sheardisc}).

\subsubsection{Comparison of shear development}\label{sheardisc}

\begin{figure}
    \centering
    \includegraphics[width=.9\linewidth]{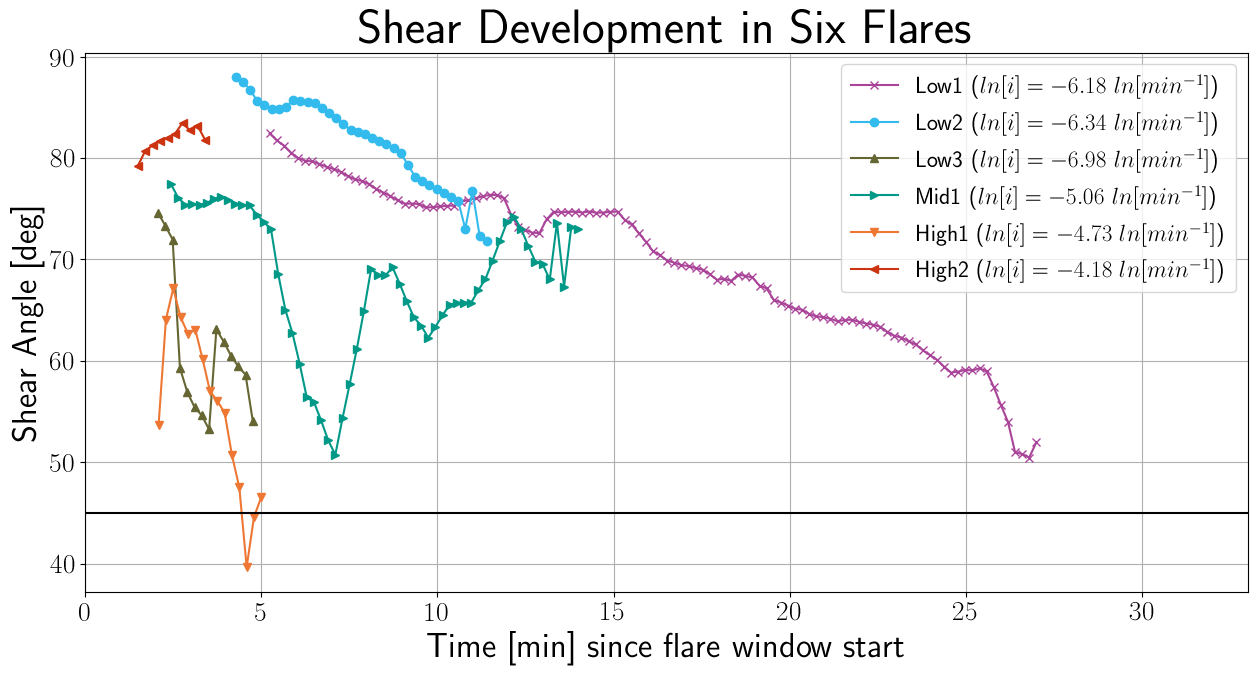}
    \caption{The evolution of shear angle relative to flare start time for the six case studies described in the above figures.  The guide field ratio is converted into a shear angle here, where larger angles correspond to more magnetic shear relative to the PIL. The horizontal line corresponds to $45^o$.}
    \label{fig:shearcomp}
\end{figure}

In Figure \ref{fig:shearcomp}, we show the time series of shear for each of the six case study events in order to address the possible connection between shear rate of change and impulsiveness.  We represent the shear as an angle $\theta$, defined in Equation \ref{gfrangle} relative to a line perpendicular to the PIL.  
Events Low1 and Low2 begin with large shear angles (greater than 80 degrees) which decrease slowly. We note that the shear angles approaching 90 degrees indicate magnetic field loops which are nearly parallel to the PIL early in flare evolution. 
 Very high shear angles have been both observed in other flare studies \citep[e.g][]{qiu_2023} and simulations \citep[e.g.][]{dahlin2021}.  Event Low2 has a minimum shear angle of roughly 72 degrees, while shear in event Low1 decreases to 50 degrees by the end of the window of observations, about 27 minutes after the start of the event.  Event Mid1 experiences a sharp decrease in shear early in the event, between 5 and 7 minutes after the start.  Event High1 begins at an even lower shear angle of roughly 60 degrees, which decreases sharply until the shear angle is roughly 40 degrees only 4 minutes after the start of the event.

In events Low1, Low2, Mid1, and High1, the events with persistently large amounts of magnetic shear also have lower impulsiveness. Event Mid1 and particularly event High1 show a signature of sharp decrease in magnetic shear, which is not present in events Low1 and Low2.   We note that when we characterize shear angle with $\theta$ rather than GFR, the decrease in shear in event High1 is more apparent, and the low shear in event Low3 relative to Low1 and Low2 is more clear.

The comparison between impulsiveness and shear is not as straightforward in events Low3 or High2.  Event Low3 begins at a moderately high shear angle of about 75 degrees, which then appears to decrease sharply.  We note that the determination of shear angle for event Low3 is affected by its short duration in the EUV lines relative to HXR.  The calculated amount of shear in the active region increases after this initial decrease, suggesting that the limited temporal scale of event Low3 introduces a source of error to our analysis.  The magnetic topology of event High2 is complex relative to other events as observed in the 171 \AA\ line, which casts doubt regarding our determination of the magnitude and rate of change of magnetic shear for this event.

\subsubsection{Application to flare energy release}

According to \cite{dahlin2021}, \cite{arnold2021}, and \cite{qiu2022}, a high amount of shear, especially early in a flare, typically delays the particle precipitation which causes the release of chromospheric emission.  \cite{dahlin2021} demonstrate the anti-correlation between particle precipitation and shear through simulation, while \cite{qiu2022} do so through observations. We remind the reader of the potential connection between shear and rates of energy deposition in the chromosphere: since the precipitation of particles heats the lower atmosphere and leads to the emission associated with flare ribbons, high early rates of particle precipitation should produce a higher impulsiveness event. If the results of these simulations are applied to observations of solar flares, we expect to also see an anti-correlation between magnetic shear and impulsiveness in our case study analysis.  

We verify in events Low1 and Low2 that there is delay in the onset of HXR emission when there is a large amount of shear in an active region (panels (a), (b), (d), and (e) of Figure \ref{fig:low1low2}) and that this delay is shorter in event Mid1, when the GFR declines quickly.  From Figure \ref{fig:shearcomp}, in four of the six events, a higher impulsiveness corresponds to a lower, less persistent amount of magnetic shear, and vice versa.  The presence of a highly sheared magnetic field may also explain the lack of HXR emission and relatively low peak flux in EUV lines in event High2.  

Of course, it is noteworthy that our analysis of this pattern is limited to only six events.  To determine whether the relationship observed between shear and the timing of flux development in events Low1, Low2, Mid1, and High1 applies to high-impulsiveness events in general, it would be necessary to identify high impulsiveness events which both (1) show a high peak flux, allowing for longer flare durations, and (2) do not suffer from saturation effects, as X-class flares observed by SDO/AIA in UV/EUV wavelengths often do \citep{kazachenko2017}. 

Even with an extensive study of shear development including many solar flares, the variation in flare reconnection and ribbon evolution complicates the connection between inferred magnetic configuration and energy release.  For example, \cite{aulanier_dudik_2019} and \cite{zemanova_2019} have observed and presented a model for flares with a different reconnection pattern.  In this model, reconnection of the field of the rising flux rope with the surrounding field produces flare ribbon motion which deviates from the strong-to-weak shear pattern.  In our case study analysis, we used AIA 171 \AA\ observations, accessed via JHelioViewer \citep[][not pictured]{muller2017}, to verify the patterns in shear evolution which we report.  For example, from the SDO/AIA 171 \AA\ images, the coronal loops during flare High2 appear to connect opposite ends of the active region, and change very little over the course of the flare.  We found that the calculated magnitude of magnetic shear based on flare ribbon position and geometry for this event changes very little.  Could this be an example of a phenomenon similar to that presented by \cite{aulanier_dudik_2019} and described in \cite{kazachenko2022}, wherein the feet of an erupting flux rope are observed as hooked extensions on opposite edges of observed flare ribbons? Whether or not this is the case, the coronal loop structure in event High2 does not adhere to the fundamental model of strong-to-weak shear evolution.  As with many aspects of the flaring process, the patterns observed here between impulsiveness and the evolution of shear are not expected to occur in all solar flares.

\section{Conclusions}\label{sec:conc}

In this study, we have sought to better understand the rate and timing of chromospheric energy release in the HXR and EUV of a large sample of solar flares relative to associated chromospheric ribbon motion.  We have analyzed Sun-as-a-star light curves, chromospheric full-disk images of the Sun, and measurements of the line-of-sight magnetic field.  The main results of our work are as follows:

\begin{enumerate}
    \item We have calculated the impulsiveness index ($i = [(I_{max}-I_{sq})/I_{sq}]/t_{1/2}$) for N = 1368 flares in the \verb+RibbonDB+ catalog from 30 April 2010 to 26 May 2014. This classification system reflects the nature timing and flare energetics of the impulsive phase of a solar flare in He {\tiny II} 304 \AA, corresponding to the period of high-energy collision of particles precipitating from the reconnection region with the chromosphere, and does not supersede other classifications of solar flares, such as according to GOES class or association with an eruption, which capture other flare properties.
    \item The impulsiveness index does not have a strong linear relationship with the SDO/EVE 304 \AA\ Sun-as-a-star light curve rise phase duration, decay phase duration, overall flare duration, or peak GOES SXR flux.  The impulsiveness index is moderately correlated with peak reconnection rate, and  flares of high peak reconnection rate belong disproportionately to the high impulsiveness category.
    \item We apply a semi-autonomous procedure for determination of PIL-relative flare ribbon motion, reconnection electric field, and magnetic shear to a selection of six case study flares, with three low impulsiveness, one mid-impulsiveness, and two high-impulsiveness events.
    \item We verify that, for the low impulsiveness events (Low1, Low2, and Low3) and mid-impulsiveness event (Mid1), PIL-parallel ribbon motion occurs mostly during the rise phase of a flare as observed in the EUV (304 \AA\ and 1600 \AA) or HXR light curves, whereas PIL-perpendicular ribbon motion occurs throughout the duration of the flare, in both the rise and decay phases. The high impulsiveness events (High1, High2) do not adhere strictly to this pattern.  The short timescales of High1 and High2 may impact our ability to interpret these data.
    \item The reconnection electric field strength is generally strongest during the rise phase and declines after the peak in chromospheric emission.
    \item We first characterize shear through the guide field ratio, the strength of the guide field component relative to the reconnecting field.  In events Low1, Low2, Low3, Mid1, and perhaps also High1, we find that HXR emission and EUV emission peak after the amount of magnetic shear in the magnetic field topology of a region declines. 
    \item In events Low1 and Low2, shear as measured by the angle $\theta$ begins high and declines slowly during the impulsive flare phase. The mid-impulsiveness event Mid1 begins with a lower shear angle which decreases sharply. The high impulsiveness event High1 begins with an even lower shear angle which also decreases sharply.  Events Low3 and High2 depart from this pattern, as discussed in Section \ref{sheardisc}.
\end{enumerate}

Items 5, 6, and 7 in the list above are evocative of recent work relating magnetic configuration to flare energy release rates.  \cite{dahlin2021}, \cite{arnold2021}, \cite{daldorff2022}, and \cite{qiu2022} indicate an inverse relationship between rates and timing of particle precipitation and amounts of magnetic shear. This phenomenon has its roots in the relatively inefficient particle energization provided by the super-Dreicer reconnection electric field acceleration when compared to Fermi acceleration via curvature drift.  The relationship between shear and particle acceleration is well-established through simulations \citep{dahlin2016,arnold2021,dahlin2021}. Based on a standard model for flare reconnection extended to three dimensions, the suppression of particle acceleration (measured here by the timing of Fermi HXR emission) should extend to energy release in UV/EUV, as the chromospheric condensation which excites the lines responsible for this emission is due to particle precipitation from the reconnection region.  

However, the non-potential field indicated by a high amount of shear early in the development of a flare is likely responsible for the storage of free energy which contributes to the flare.  Therefore, the unshearing process is essential to the development of a solar flare, as it requires some amount of shear (non-potentiality) early in the flare.  The findings of \cite{dahlin2016}, \cite{arnold2021}, and \cite{dahlin2021} may explain why lower impulsiveness case studies show a relatively high, persistent amount of magnetic shear. Given the nature of our work here, we can not expand this discussion to flares in general.  However, application of simulations and comparison to observation may substantiate this potential connection.  

The three-dimensional simulation outputs from \cite{dahlin2021} may offer the opportunity to study the effect of shear variations on flare development.  \cite{dahlin2021} utilize the Adaptively Refined Magnetohydrodynamics Solver \citep{devore_antiochos_2008}, which injects energy into an idealized active region with two sets of dipoles superimposed on a background solar dipole. The injected energy takes the form of shear flux, which is added to the system until an eruption occurs via breakout.  The resulting simulation can be used to track the evolution of chromospheric ribbon fronts, magnetic field lines, current-density magnitude, and CME release when there is an associated eruption. To test whether there exists a relationship between impulsiveness and shear, if it is possible to simulate eruptions with a range of initial shear values prior to eruption, production of many simulated flare light curves and calculation of the impulsiveness index associated with the resulting eruptions could advance our understanding of how impulsiveness, ribbon geometry, and magnetic field topology relate.  There are several complications in applying ARMS in this way.  The simulation domain of ARMS as produced by e.g. \cite{dahlin2021} is much larger than a typical active region, making a comparison to observed flare intensities a concern.  Were we to apply a series of ARMS simulations to our work, it would be necessary to determine the relationship between impulsiveness and, for example, reconnection rate.  As we have shown, there is a weak relationship between impulsiveness in the 304 \AA\ line and reconnection rate.  In order to adequately apply ARMS to a discussion of impulsiveness and shear, it is therefore useful to first study the variations in impulsiveness (particularly as related to other flare quantities) in other wavelengths.  

Alternatively, the coronal extension of the three-dimensional MURaM radiative MHD code \citep{rempel_2017} has recently been used to produce different setups with varying rates of magnetic free energy injection and different evolution patterns of the sheared magnetic arcade (SMA) or magnetic flux rope (MFR) \citep{rempel_2023}.  Both setups show chromospheric flare ribbons.  The flare simulations of \citep{rempel_2023} reach GOES class M1, meaning that the updated MURaM simulations produce flares of comparable size to those discussed in our case study analysis.  The MURaM code synthesizes EUV emission for different AIA channels. The versatility of these MURaM simulations and synthesis of observables similar to those used in our work could be used to track the calculated evolution of magnetic shear while also determining impulsiveness for a variety of flares. As a caveat, we must be cautious in interpreting MURaM's synthesis of non-local thermodynamic equilibrium (LTE) lines such as 304 \AA, which originates partially in the transition region and is therefore affected by large gradients in, for example, temperature and density. The version of MURaM presented by \cite{rempel_2017} assumes LTE.

 In the future, comparison of the impulsiveness index for solar flares to spectral signatures and other flare properties studied in stellar flares would provide a meaningful link between solar and stellar physics.  With the development of an impulsiveness index with the same formulation as that developed for flares from M dwarf stars in \cite{kowalski2013}, we could make a direct comparison between solar and stellar impulsive events.  \cite{kowalski2013} found that impulsiveness has a relationship with several optical spectral properties in stellar flare observations. Direct application of this analysis to solar flares may allow us to determine if and why these relationships hold for our star.  The application of the impulsiveness index to solar flares and the relationship between impulsiveness and other quantities may provide insight into stellar flare evolution which would not be possible without solar observations.  Such an approach is similar to that of \cite{namekata_2017}, who compare a sample of white light solar flares to superflares on solar-type stars and argue that the scaling law reconciling one discrepancy between flare duration and energy could be used to predict stellar parameters outside the resolution-limited realm of observation.  To overcome the limits of stellar flare spectral and spatial resolution, it will be valuable to compare the distribution of impulsiveness for solar flares to that of stellar flares.

However, we must be conscious of the limitations currently involved in the application of the impulsiveness index to solar flares.  The statistical study of impulsiveness presented in Section \ref{sec:statstudy} is limited to flares observed by MEGS-A between 2010 April 30 and 2014 May 26 and to flares with sufficient S/N to confidently resolve a flare signature. We include both eruptive and confined flares in our sample and have not investigated the potential difference in the distribution of impulsiveness values between these two populations.  We include GOES class C, M, and X flares in our sample, but do not find a relationship between impulsiveness and SXR flux.  In fact, besides the observation that flares with high reconnection rate tend to have high impulsiveness values, we find very little correlation between impulsiveness and other flare properties.  While it may be the case that impulsiveness measures the energy influx into the chromosphere in a manner not reflected in the other flare properties studied here, these results also may support the conclusion that the flare process is too complex to be meaningfully characterized by impulsiveness.  Investigation of other flare properties, such as the spectral signatures studied by \cite{kowalski2013}, may resolve this issue.

The selection of the 304 \AA\ line, which captures emission from chromospheric flare ribbons in addition to some filling of flare loops, results in an impulsiveness index with specific idiosyncrasies.  Any line or continuum flux used to develop an impulsiveness index will represent different heating and cooling physics and therefore present different values for impulsiveness.  Figures 14-17 in \cite{kowalski2013} show significantly different light curve development depending on the line or continuum flux studies, even if these curves result from similar physics. Comparison of the C4170 (continuum at 4170 \AA) light curve to the $H\alpha$ light curve (which we might expect to show similar evolution to He {\tiny II} 304 \AA) shows a longer cooling time in the latter, though the rise phase evolution is similar. 

Most of our case study events display HXR emission for roughly the time period covered by $t_{1/2}$ in the 304 \AA\ light curve, but in event High1 there is significant 304 \AA\ emission after HXR emission has essentially ceased due to other, potentially non-chromospheric, sources of 304 \AA\ emission.  This indicates that $t_{1/2}$ in 304 \AA\ is not always a reliable proxy for the traditionally-defined impulsive phase, as the 304 \AA\ line clearly does not always solely capture emission due to particle precipitation.  As a result, one of our main findings is that the impulsiveness of an event will vary depending on the wavelength or bandpass from which it is calculated.  While the use of SDO/EVE 304 \AA\ data is an appropriate initial choice for our work here as described in Section \ref{sec:datainstr}, we must use caution in comparing our distribution of impulsiveness values to the impulsiveness as measured with flare light curves other than those of 304 \AA. Here, we can only confidently say that the calculated ``impulsiveness" values capture the evolution of the 304 \AA\ line itself. A study that calculates impulsiveness for other lines or bands may reveal that there exists an option for the development of the impulsiveness index that is more directly reflective of the traditionally-defined impulsive phase.  For example, a recent study by \cite{jing2024} reported on a sample of white-light flares between October 2022 and May 2023 observed in the Balmer continuum at 360 nm by the Advanced Space-based Solar Observatory.  While the number of flares captured by this instrument (launched in October 2022) is currently limited, this is an example of a promising option for determination of an impulsiveness index more useful in explaining the empirical relations between stellar flare impulsiveness and certain spectral quantities.

Finally, some quantification of shear magnitude and intensity of shear decrease for all events in our sample of 480 flares would provide valuable observational insight into the predictions suggested by a connection between magnetic field topology and energy release in the lower solar atmosphere. By expanding our case study analysis to a much larger sample, we may be able to determine whether there is a general relationship between magnitude and evolution of magnetic shear and spatially integrated light curve characteristics.

\hfill \break

We would like to thank the anonymous referees for their helpful comments, which notably aided in improving several aspects of our study. Support for this work is provided by the National Science Foundation through the DKIST Ambassadors program, administered by the National Solar Observatory and the Association of Universities for Research in Astronomy, Inc. and the George Ellery Hale Graduate Fellowship. M.D.K. acknowledges NASA grant ECIP NNH18ZDA001N, and NSF CAREER grant SPVKK1RC2MZ3. We are grateful to Dr. Joel Dahlin, Dr. Bradley Hindman, Dr. Ann-Marie Madigan, Dr. Lauren Blum, Dr. Ryan French, and Dr. Yuta Notsu for valuable discussions regarding the physical basis and broader impact of this work. 

\clearpage
\appendix
\section{Details of the statistical study}\label{sec:appendix}
In the following, we provide details of the statistical study of impulsiveness, including, in Appendix \ref{lcanalysis}, the selection of events for comparison to other flare parameters, and in Appendix \ref{sec:modeling}, a discussion of the modeled impulsiveness distributions after event vetting and selection.

\subsection{Selection of events}\label{lcanalysis}

From the 2049 events during which MEGS-A was operational, we use SDO/EVE 304 \AA\ light curves and SXR start times to identify the precise start, peak, and end times (indicated in Figure \ref{examplecurve}) for each flare in the 304 \AA\ line.  The start times for each flare as identified in the EUV or SXR are often offset significantly.  We therefore developed a procedure for flare detection to accurately identify temporal flare parameters in the 304 \AA\ line.  The procedure identifies flares by determining when the flare light curve has exceeded solar quiet flux values by two standard deviations from the estimated solar quiet, for a variable number of points (between 10 and 40 points).  The procedure begins with strict requirements for flare detection parameters (i.e. a larger of points exceeding solar quiet) and relaxes these requirements incrementally until a flare is found in the window defined by \verb+RibbonDB+.  We determine the minimum and maximum stringency of these parameters by inspection.  

In the decay phase of the flare, our method tracks the number of time steps for which the irradiance value has been less than half of one standard deviation from the estimated solar quiet.  The end time of the flare is the point at which the number of time steps for which this is the case reaches 50.  This criterion is relaxed for some flares, based on inspection.  The method intentionally results in a conservative estimate of flare end time, since, for the determination of the impulsiveness index, we require only the temporal full width at half height of the light curve.  Defining the end of the flare in this way ensures that we consistently capture the entire evolution of the flare. 

When we are unable to determine start, peak, and end times of a flare included in the \verb+RibbonDB+ catalog using the 304 \AA\ light curve, we exclude it from our analysis.  In some cases, temporal flare parameters are not identified due to low signal-to-noise ratio (S/N). In others, the rise phase of a flare occurs while there is still significant emission from a previous event, making it difficult to confidently identify the end time of the first event and the start time of the second.  Occasionally we can not reliably determine the parameters $I_{max}$, $I_{sq}$, and $t_{1/2}$ from which impulsiveness is derived, and therefore we do not consider the flare for analysis.  However, our method does not necessarily exclude partially overlapping flares.  If the first event has returned sufficiently close to solar quiet prior to the rise phase of the second event, we find that the determination of flare parameters and calculation of the impulsiveness index for both events are reliable.  However, if a second event begins too early in the decay phase of the first, it is not possible to reliably calculate impulsiveness, and both events are excluded from our analysis.  Out of the original 2049 flares in the intersection between \verb+RibbonDB+ and the coverage of MEGS-A, the result is a sample of $N=1368$ flares which show clear signatures in the 304 \AA\ line with definite rise and decay phases.

Here we describe the selection of a subsample of events.  The sample is large enough to include all impulsiveness categories with sufficient representation, but small enough to be confident that each impulsiveness value is accurately calculated by flare processing without verification of each individual impulsiveness value.  To select events for our impulsiveness index study, we separately model the rise and decay phases of each flare in the selection and test the fidelity of the models in comparison to the data. We choose the models for the rise phase and decay phase based on statistical preference over others tested via curve fitting software in MATLAB.  We find that a two-term exponential function and two-term Fourier series best fit the rise phase and decay phase, respectively, of each flare.  The two-term Fourier series model accounts for the occasionally observed EUV late-phase flaring well \citep{woods2014}.  

The flares with a reliably high signal-to-noise ratio fit both models well.  To identify a subset of events for which we are most confident in the accuracy of our impulsiveness determination, we generate a subset of flares with rise and decay phase data which fit the two-term exponential and two-term Fourier series models best, respectively.  We consider the Pearson's correlation coefficients between model and data, $r^2_{rise}$ and $r^2_{decay}$, for the rise and decay phases, respectively.

From the rise and decay phase model-data correlation coefficients, we calculate the score $s = r^2_{rise}\cdot r^2_{decay}$ to measure the reliability of collected data and parameters for each flare.  For statistical comparison to other flare properties, we select the 480 best-performing events with values of $s = 0.81$ or above, which corresponds to an average $r^2$ of $0.90$ for both models and no lower than $r^2 = 0.81$ for either individual model-data comparison, separately.

\begin{figure}
\begin{center}
   \includegraphics[width=.75\linewidth]{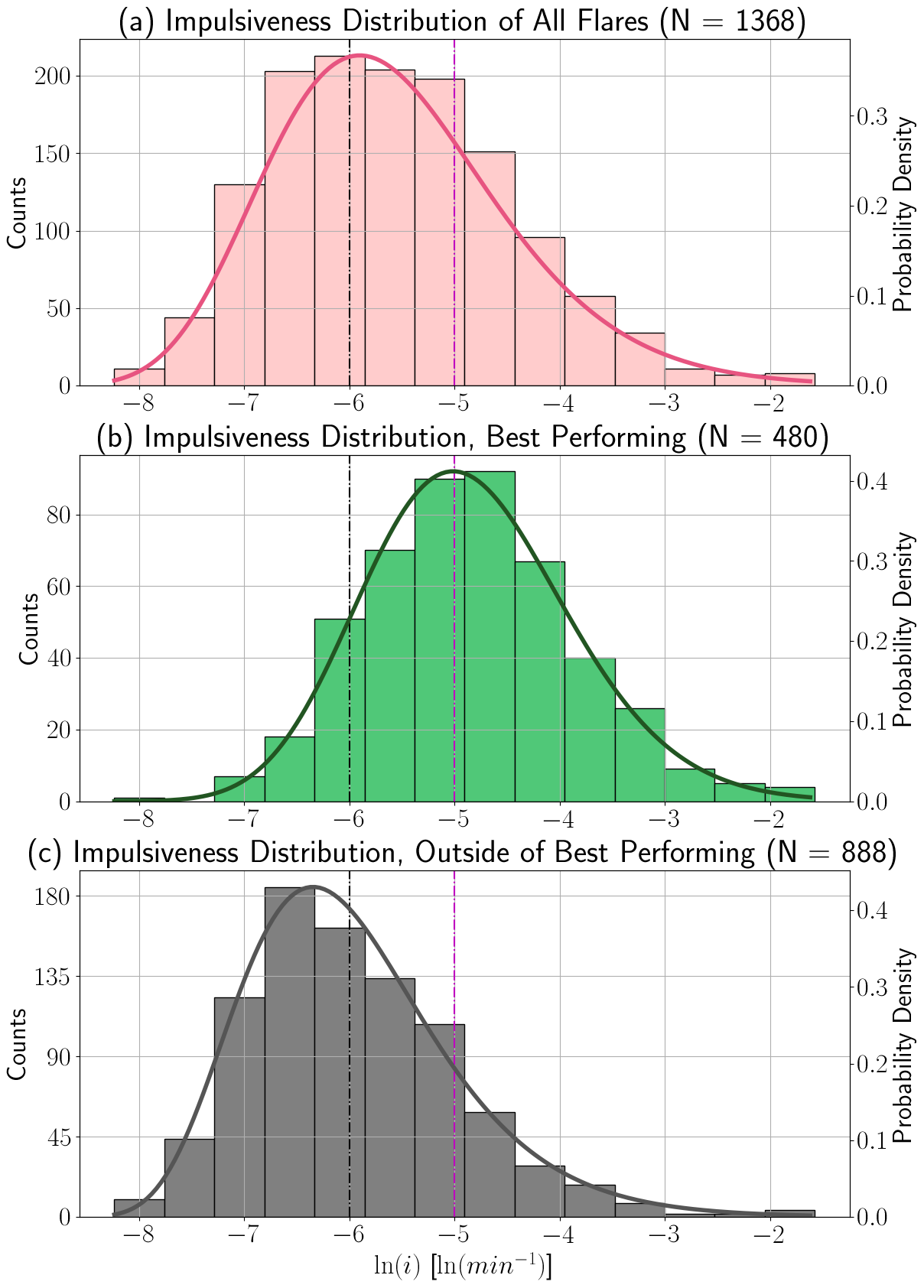}
  \caption{Distributions for impulsiveness determined from the 304 \AA\ light curves for flares in the \texttt{RibbonDB} catalog captured by SDO/EVE MEGS-A.  Panel (a) represents all flares that we selected from the \texttt{RibbonDB} catalog. Panel (b) represents the flares used for comparison to other quantities, selected according to their compliance with models for the rise and decay phases.  Panel (c) represents the flares in the sample not included in the best-performing 480 flares.  The black vertical dash-dotted line indicates the high cutoff for low-impulsiveness events. The violet vertical dash-dotted line indicates the low cutoff for high impulsiveness events. Probability density functions derived from higher-resolution impulsiveness binning (as described in Section \ref{sec:modeling}) are overlaid in each panel.}
  \label{compsamp}
  \end{center}
\end{figure}

Figure \ref{compsamp} shows the distributions of all flares, the best-performing 480 with high scores $s$, and the flares outside the best-performing 480.  These distributions are discussed at length in Section \ref{sec:modeling}.

\subsection{Distribution of flare impulsiveness indices}\label{sec:modeling}

Figure \ref{compsamp} shows impulsiveness distributions and models fit to these distributions.  To fit the impulsiveness distributions, we use Python's \texttt{distfit} package and compare several possible models for the distribution function.  The \texttt{distfit} package uses residual sum of squares (RSS) to measure the amount of variance of the data around a given model.  We find that the preferred model for the three distributions of all flares, best-performing flares, and flares outside the best-performing is the log-normal distribution, defined as

\begin{equation}\label{lognorm}
f(x,s) = \frac{a}{\sigma(x-x_0)\sqrt{2\pi}}\exp\left(-\frac{\ln^2[(x-x_0)/a]}{2\sigma^2}\right),
\end{equation}

\noindent where $\sigma$ is the standard deviation of the distribution's natural logarithm, $x_0$ is the location parameter, which measures the shift of the distribution, and $a$ is the scale parameter, the expected value of the distribution's natural logarithm.  Table \ref{tab:parameters} shows the fit parameters for samples of all flares, the best-performing flares, and flares outside the best-performing.
\!
\begin{deluxetable}{|c|c|c|c|c|}
\tabletypesize{\footnotesize}
\tablewidth{0pt}
 \tablecaption{Log-normal fit parameters for all, best-performing, and outside of best-performing flare samples.  $N$ is the number of flares in the sample; $\sigma$ is a measure of the spread of the distribution; $x_0$ is the location parameter, which measures the shift of the distribution; and $a$ is the scale parameter.}
 \label{tab:parameters}
 \tablehead{
 \colhead{Sample} & \colhead{$N$} & \colhead{$\sigma$} & \colhead{$x_0$} & \colhead{$a$}} 
\startdata
 All & 1368 & 0.20 & -11.21 & 5.52 \\
    \hline
Best-Performing & 480 & 0.11 & -13.57 & 8.66 \\
    \hline
 Outside Best-Performing & 888 & 0.27 & -9.65 & 3.55 \\
    \hline
\enddata
 \vspace{-0.5cm}
 
\end{deluxetable}

We generate model log-normal PDFs for all three distributions at a resolution roughly 10 times higher than the bin sizes in the histogram of Figure \ref{compsamp}.  Figure \ref{compsamp} shows the resulting PDFs.  We compare the sample of best-performing flares to the sample of flares outside the best-performing using a two-sample Kolmogorov-Smirnov (K-S) test from \texttt{scipy.stats.ks\_2samp} to determine whether the method by which we select flares according to their adherence to models of the rise and decay phases has selected for higher impulsiveness.  The test returns a K-S statistic of 0.441 and a p-value which is practically zero and well below the 5$\%$ significance level, providing evidence to suggest that we may reject the null hypothesis that the two samples are drawn from the same distribution.  We also perform a \textit{t}-test for difference in means on the best-performing flares and the flares outside the best-performing.  The \textit{t}-test results in a p-value near zero, suggesting to a high degree of statistical significance that the two samples have different means.

Given the results of the hypothesis tests, it is reasonable to conclude that our method of flare selection for statistical analysis is biased towards higher impulsiveness.  This is not an unexpected result, as flares with a higher peak irradiance relative to the solar quiet (which will tend to produce a higher impulsiveness index) typically exhibit more well-defined light curves relative to the background, and are therefore more easily discerned by the model-fitting procedure.

\clearpage
\bibliographystyle{aasjournal}
\bibliography{flare_impulsiveness_manuscript}

\end{document}